\pdfoutput=1
\documentclass[12pt]{iopart}
\usepackage{iopams}
\usepackage{graphicx} 
\usepackage{subfigure}
\usepackage{ifpdf}
\usepackage{booktabs}
\usepackage{cases}
\usepackage{color}
\usepackage[utf8]{inputenc}
\usepackage[T1]{fontenc}
\newcommand{\ie}{i.e., }
\newcommand{\eg}{e.g., }
\ifpdf
    \usepackage{microtype}
\else

\fi

\begin{document}

\title[The comfortable driving model revisited]{The comfortable driving model revisited: Traffic phases and phase transitions}

\author{Florian Knorr and Michael Schreckenberg}

\address{Fakult\"at f\"ur Physik, Universit\"at Duisburg-Essen, 47048 Duisburg, Germany}
\eads{\mailto{florian.knorr@uni-due.de}, \mailto{michael.schreckenberg@uni-due.de}}

\begin{abstract}
We study the spatiotemporal patterns resulting from different boundary conditions 
for a microscopic traffic model and contrast them with empirical results.
By evaluating the time series of local measurements, the local traffic states 
are assigned to the different traffic phases of Kerner's three-phase traffic theory.
For this classification we use the rule-based FOTO-method, which provides `hard' rules 
for this assignment. 
Using this approach, our analysis shows that the model is indeed able to reproduce 
three qualitatively different traffic phases: free flow (F), synchronized traffic (S), and wide moving jams (J).
In addition, we investigate the likelihood of transitions between 
the three traffic phases. 
We show that a transition from free flow (F) to a wide moving jam (J) often 
involves an intermediate transition; first from free flow F to synchronized flow S and then from synchronized 
flow to a wide moving jam.
This is supported by the fact that the so-called F$\rightarrow$S transition (from free flow to synchronized traffic) 
is much more likely than a direct F$\rightarrow$J transition. 

The model under consideration has a functional relationship between traffic flow and traffic density. 
The fundamental hypothesis of the three-phase traffic theory, however, postulates that the steady states 
of synchronized flow occupy a two-dimensional region in the flow-density plane. 
Due to the obvious discrepancy between the model investigated here and the postulate of the three-phase 
traffic theory, the good agreement that we found could not be expected.
For a more detailed analysis, we also studied vehicle dynamics at a
microscopic level and provide a comparison of real detector data with simulated data of the 
identical highway segment.
\end{abstract}

\maketitle

\section{Introduction}
Assessing a model's quality is probably best done by comparing its results with empirical findings. 
For microscopic traffic models, this approach is particularly interesting because different scales have 
to be considered. On the one hand, microscopic traffic models are  based 
on the assumption that a vehicle's motion is governed by the next (or the next two) 
vehicles ahead. Hence, from a single vehicle's perspective it makes no difference 
in these models whether there are only two or two thousand vehicles on the road 
(for a review of microscopic traffic models from a physical point of view, see~\cite{Helbing2001,SchadschneiderChowdhuryNishinari2010}).
Many empirical studies of traffic, however, focus on macroscopic phenomena. 
Emerging behaviors (\eg traffic breakdown, jam formation), for instance, require the coordinated 
motion of hundreds of vehicles (see the recent review article~\cite{BellomoDogbe2011}, which also discusses current challenges on traffic modeling). 

To characterize traffic flow, the distinction between freely flowing and congested traffic 
is obvious, but also quite coarse-grained. A more detailed analysis of traffic reveals a rich 
variety of spatiotemporal patterns in congested traffic: 
the meticulous study of empirical traffic data has led to the development of the three-phase theory 
of traffic (exhaustively presented in the books by Kerner \cite{Kerner2004,Kerner2009}). 
According to this theory, congested traffic subdivides into two phases: `synchronized traffic' (S) and 
`wide moving jams' (J). 
Low average velocities, low vehicle flow rates, and a downstream front that propagates with a constant velocity 
against the vehicles' direction of travel are characteristic features of the latter. 
In contrast, the downstream front of synchronized 
traffic is often located at a bottleneck and, although average velocities are considerably below the 
velocities of free flow, traffic flow is higher than in jammed traffic---sometimes close to the rates observed in free flow.

Kerner's detailed analysis has led to several conclusions about the characteristics of traffic flow. 
One of these is the \textit{fundamental hypothesis of three-phase traffic theory}. 
It states that ``steady states of the synchronized phase cover a two-dimensional region in the flow density-plane''~\cite[p.~46]{Kerner2009}.
From this follows that models with a functional relationship between vehicle flow and density
are not able to adequately reproduce the phases of congested traffic. 
Recently, this conclusion was discussed controversially though 
\cite{SchoenhofHelbing2007, SchoenhofHelbing2009, TreiberKestingHelbing2010}, and an alternative explanation 
for the two-dimensional region of steady states was given \cite{TreiberKestingHelbing2006}.

The distinction between these `traffic phases' gets even more complicated for several reasons: 
(i) Synchronized traffic itself subdivides into various classes with different spatiotemporal characteristics. 
(ii) Although some of these sub-classes seem to be identical to the ones found by Sch\"onhof and Helbing \cite{SchoenhofHelbing2007, SchoenhofHelbing2009}, Helbing \etal \cite{HelbingTreiberKestingSchoenhof2009}, and Treiber \etal \cite{TreiberKestingHelbing2010}, the 
just mentioned authors and Kerner~\cite{Kerner2004,Kerner2009} 
use a different vocabulary to classify these patterns. 
(iii) In addition, the identification of distinct 
traffic phases is difficult, if not impossible, based on macroscopic traffic data~\cite{Kerner2011}.

In this context, it has to be noted that the term `traffic phase' does not (necessarily) correspond 
to a phase in the physical sense.
Although traffic flow can be interpreted as driven particle system out of equilibrium~\cite{SchadschneiderChowdhuryNishinari2010}, 
which exhibits boundary-induced phase transitions~\cite{Krug1991,AntalSchuetz2000}, a traffic phase 
rather represents a characteristic spatiotemporal traffic pattern.

In this paper, we present a systematic analysis of a well-known traffic cellular automaton model 
with a functionalship relationship between traffic flow and vehicle density, the so-called 
comfortable driving model~\cite{KnospeSantenSchadschneiderSchreckenberg2000,KnospeSantenSchadschneiderSchreckenberg2002} (see~\ref{app:cdm}). 
After introducing the setup of our simulation, we reconstruct the model's phase diagram 
and investigate the underlying spatiotemporal patterns and the associated traffic phases. 
Thereby, we are able to better assess the model's ability to simulate emerging phenomena and its predictive character 
for macroscopic traffic patterns.
For the classification of traffic phases, we study locally measured data and use the patented
FOTO-method (forecasting of traffic objects, see \ref{app:foto}), which was invented by Kerner 
\etal~\cite{KernerAleksicDenneler2002,KernerRehbornAleksicHaug2004}, to distinguish the three phases of traffic. 
Despite some imperfections, the FOTO-method allows an objective classification  
of traffic states that may not only help to compare empirical with simulated data, but can also be applied to compare 
traffic models with each other.
\section{Simulations}
In the following, we study the dynamics of the comfortable driving model~\cite{KnospeSantenSchadschneiderSchreckenberg2000,
KnospeSantenSchadschneiderSchreckenberg2002}, to which we will refer as CDM from now on, with open boundary conditions.
The CDM is a cellular automaton model with extensions for anticipatory driving behavior.
Earlier investigations showed a good agreement of the model with empirical traffic 
data on a microscopic level (\eg headway 
distributions)~\cite{KnospeSantenSchadschneiderSchreckenberg2002,KnospeSantenSchadschneiderSchreckenberg2004}.
As is common for cellular automata, space is discretized in sites (of length 1.5~m) and time is discretized in (time or update) 
steps (of duration 1~s).

\subsection{Open boundaries}
\label{sec:open_boundaries}
As we use a similar simulation setup to Barlovic \etal \cite{BarlovicHuisingaSchadschneiderSchreckenberg2002}, 
we give only a brief summary of the simulation method. 
(Minor modifications were necessary because each vehicle occupies $l_\mathrm{veh}>1$ sites.) 
We consider a one-lane road of $N$ sites, on which vehicles move from left to right. 
The left boundary or entrance section consists of the leftmost $v_\mathrm{max} + l_\mathrm{veh}+1$ sites, 
where $v_\mathrm{max}$ denotes the vehicles' maximum velocity. 
Vehicles, which enter the road from the left boundary with probability $\alpha$, are inserted with velocity 
$v_\mathrm{max}$ at position $x_\mathrm{insert}=\min(v_\mathrm{max} + l_\mathrm{veh}+1, x_\mathrm{last}-v_\mathrm{max})$, 
where $x_\mathrm{last}$ is the rear position of the vehicle closest to the entrance section. 
Then, all vehicles, including the newly inserted one, move according to the CDM's rules of 
motion~\cite{KnospeSantenSchadschneiderSchreckenberg2000}. 
If a vehicle is not able to leave the entrance section, it is removed afterwards. 
This insertion strategy allows for higher inflow rates compared with
the obvious insertion strategy, which places a vehicle in the leftmost site with probability $\alpha$ if 
this site is empty~\cite{BarlovicHuisingaSchadschneiderSchreckenberg2002}.

The right boundary or exit section is modeled as follows: 
before other vehicles move forward, the rightmost site is occupied with probability $\beta$, 
and it is cleared after the vehicles have moved. 
Moreover, a vehicle is removed from the road if it will reach the rightmost site or beyond 
during the next update step by maintaining its current velocity. 

One can interpret the probability $\alpha$ as the inflow of vehicles to the road segment, and the probability $\beta$ 
determines the strength of local perturbations caused at the downstream boundary. 
These perturbation may, for instance, result from vehicles that enter the road via an on-ramp and occur randomly in front of 
vehicles on the main road. 
Consequently, high values of $\beta$ result in a low exit probability of the system. 

The following results were obtained on a road consisting of $N=5001$ sites. 
The first $2\times 10^4$ from a total of $2.5\times 10^4$ time steps were discarded in each simulation run to avoid transient behavior 
(\eg in some cases, it took several thousand time steps until a jam formed at the exit boundary reached the entrance boundary). 
Vehicles could move $v_\mathrm{max}=22$~sites per time step at most and had a length of $l_\mathrm{veh}=5$~sites.

\subsection{Results}
\label{subsec:phase_diagram}
\begin{figure}[hbtp]%
	\subfigure[]{%
		\label{subfig:phase_diagram_3d}%
		\includegraphics[width=0.47\textwidth]{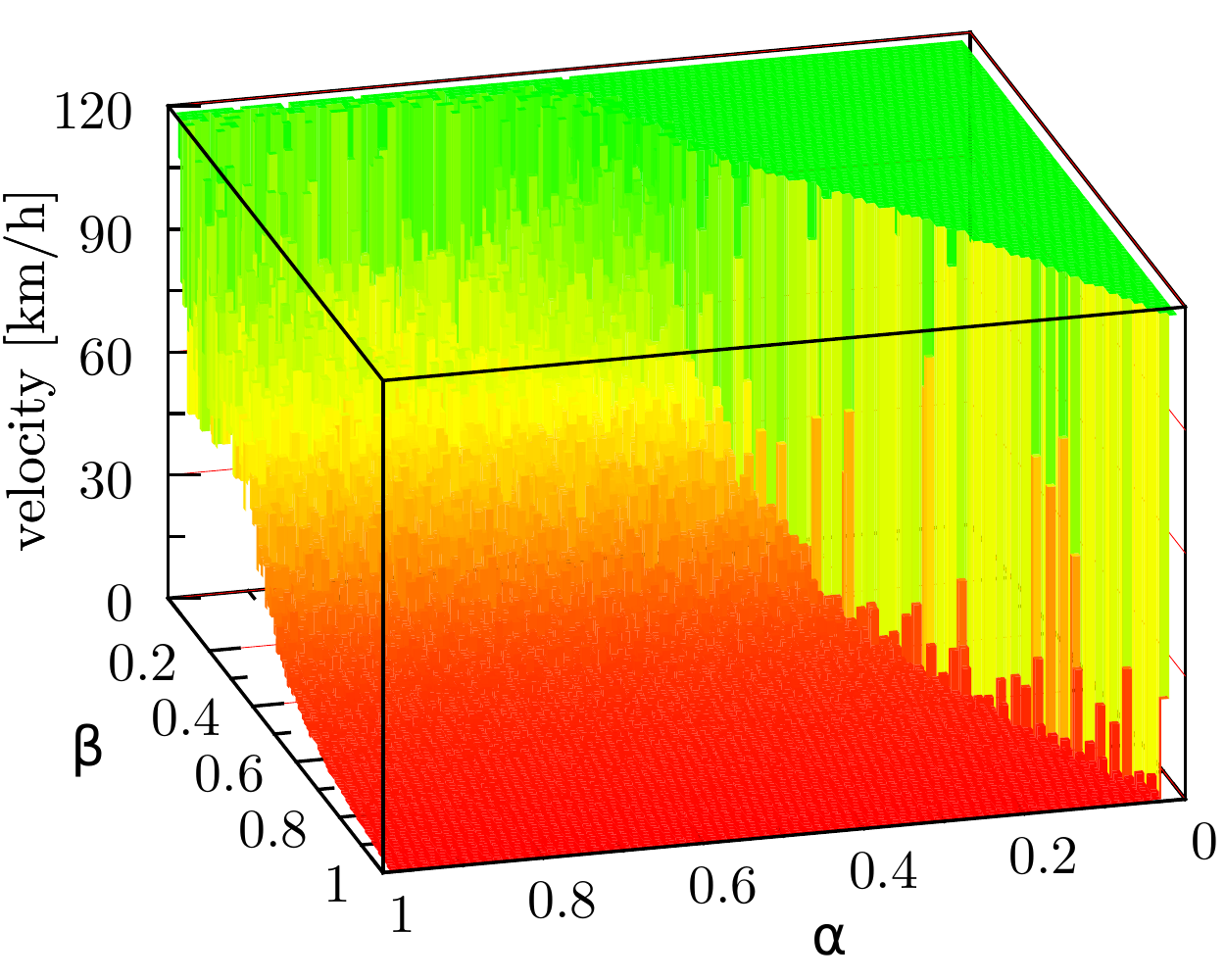}%
	}\hfill%
	\subfigure[]{%
		\label{subfig:phase_diagram_2d}%
		\includegraphics[width=0.47\textwidth]{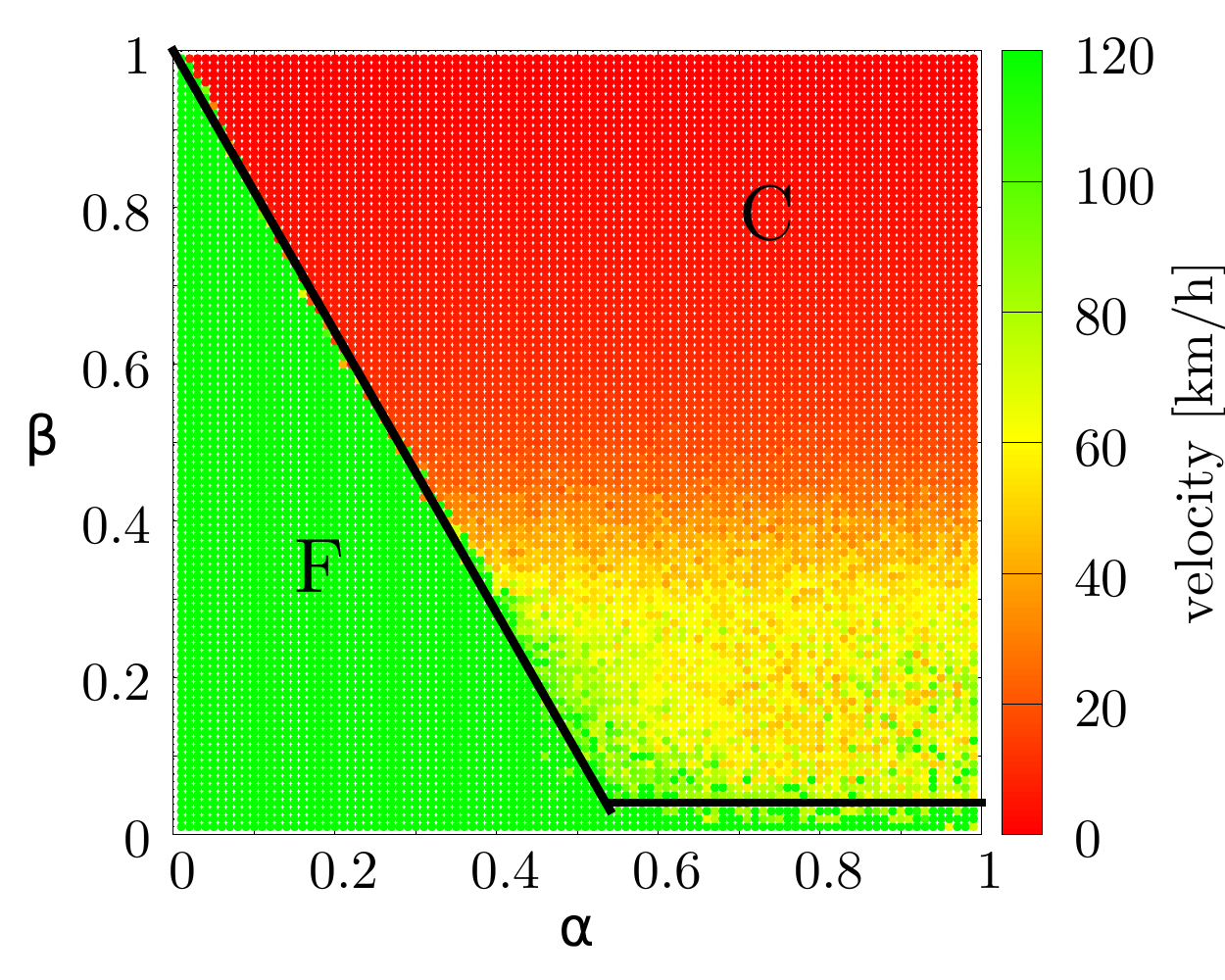}%
	}\\%
	\caption{\subref{subfig:phase_diagram_3d} The rotated 3D representation of the phase diagram 
		illustrates very well that the transition from free flow to congested traffic is discontinuous. 
		For better readability, we have added straight lines to the equivalent 2D representation in~\subref{subfig:phase_diagram_2d}, 
		which represents a stylized phase diagram. 
		(For the units of the z-axis we have used the standard conversion, where a site measures 
		$1.5$~m and an update step corresponds to 1~s~\cite{KnospeSantenSchadschneiderSchreckenberg2000}.)}
	\label{fig:phase_diagram} 
\end{figure}

We have analyzed the vehicle dynamics for all combinations of entrance and exit 
boundary conditions resulting from a step size of $0.01$ (\ie $\alpha, \beta \in [0.01,0.02,\ldots,0.99]$). 
The physical phase diagram resulting from these measurements in the bulk of the system, which
excludes the first and the last N/3 sites of the road, is depicted in figure~\ref{fig:phase_diagram}.  

The lines separating free flow (F) and congested traffic (C) in figure~\ref{subfig:phase_diagram_2d} were determined from the average 
velocity in the bulk: if the average velocity was at least 99.5\% of the vehicles' maximum velocity $v_\mathrm{max}$, free flow was assumed. 
(The value for the threshold follows from the probability $p_d$ (see \ref{app:cdm}), 
with which vehicles randomly reduce their velocity in free flow.)
This classification is probably more intuitive than the application of the extremal principle~\cite{PopkovSchuetz1999}, 
which was used by Appert and Santen \cite{AppertSanten2001} or Barlovic~\etal \cite{BarlovicHuisingaSchadschneiderSchreckenberg2002} 
for simpler models.
Since the application of the extremal principle requires the knowledge of the vehicle densities induced at the boundaries, 
this approach relies on extensive simulations as well. 
(Only for very simple models such as the TASEP~\cite{SchadschneiderChowdhuryNishinari2010}, are the boundary densities identical to 
the entrance and exit probabilities $\alpha$ and $\beta$.)
We admit, however, that our very simplistic approach of constructing the phase diagram might not be suited 
for models with a more complex fundamental diagram (\eg with local maxima of the flow~\cite{PopkovSchuetz1999}). 

As one can see, especially from figure~\ref{subfig:phase_diagram_3d}, free flow (F) and congested traffic (C) are separated 
by a very sharp, discontinuous line for $\beta \gtrsim 0.1$.
In free flow (F) practically all vehicles move at their maximum velocity. 
The averaging process, which led to the above figures, hides spatial and temporal information.
Kerner, however, defines traffic phases as ``a set of traffic states considered in space and time that exhibit some specific spatiotemporal features''~\cite{Kerner2004}.
Therefore, it is necessary to investigate the microscopic dynamics of vehicles in more detail. 

\begin{figure}[htb]
	\subfigure[]{
		\label{subfig:high_flow}
		\includegraphics[width=0.49\textwidth]{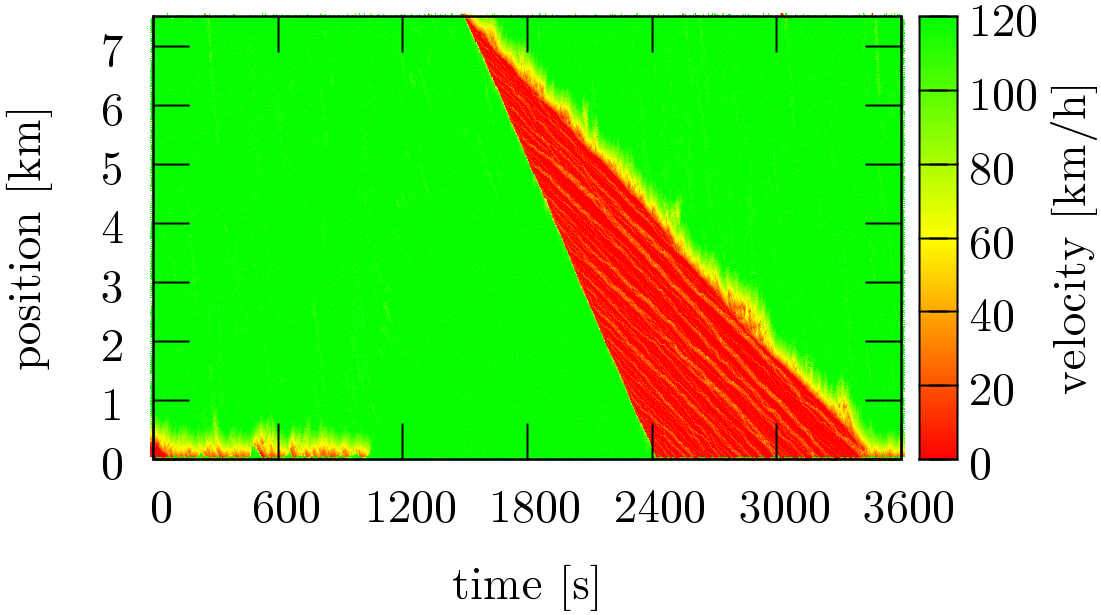}
	}\hfill
	\subfigure[]{
		\label{subfig:jam_waves}
		\includegraphics[width=0.49\textwidth]{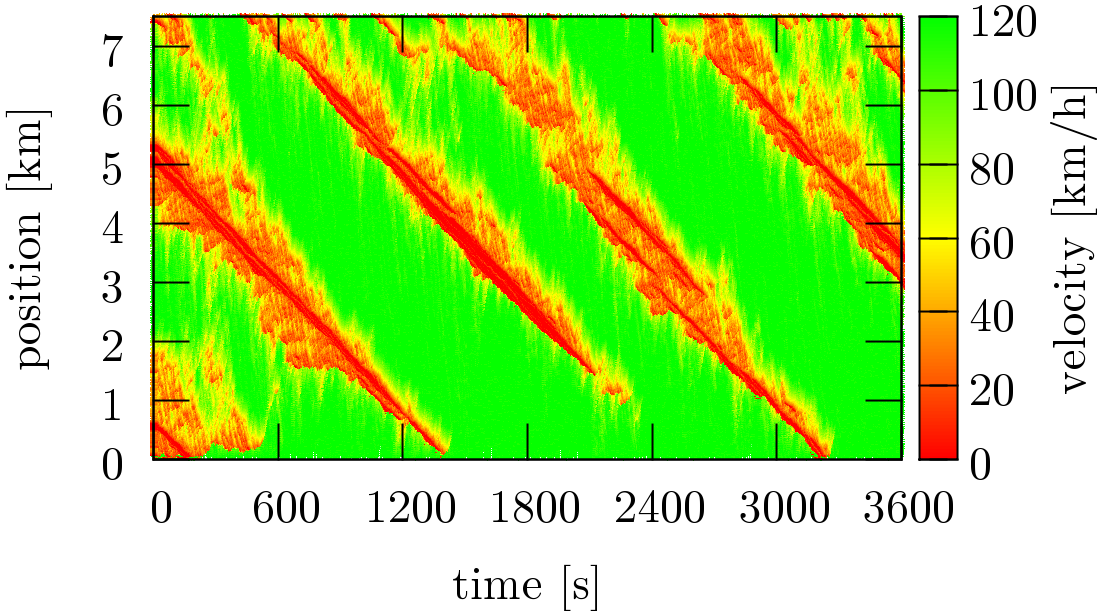}
	}\\
	\subfigure[]{
		\label{subfig:localized}
		\includegraphics[width=0.49\textwidth]{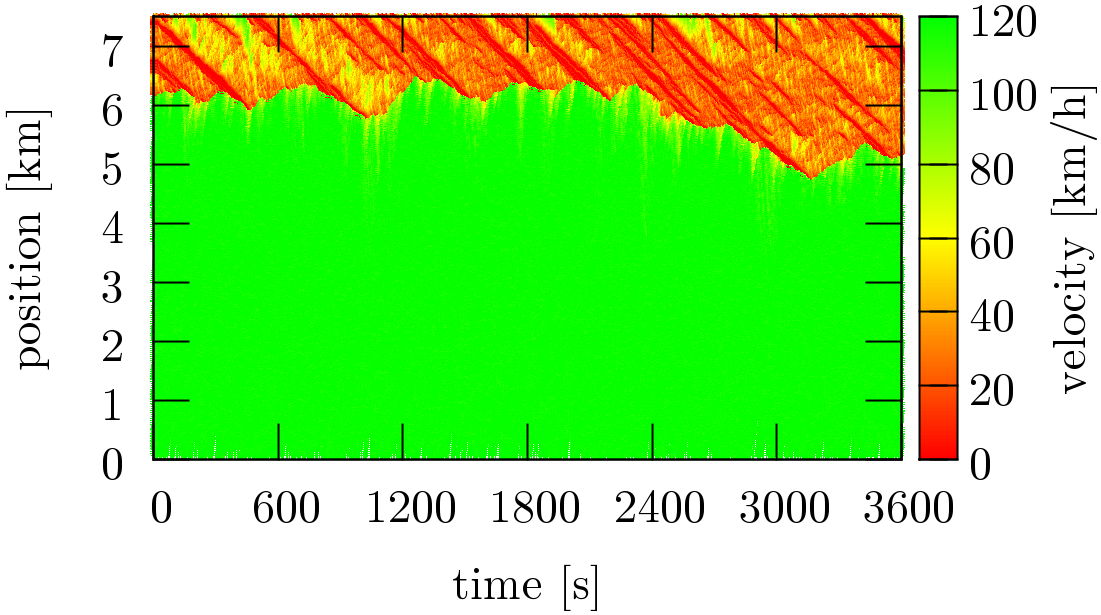}
	}\hfill
	\subfigure[]{
		\label{subfig:synchronized}
		\includegraphics[width=0.49\textwidth]{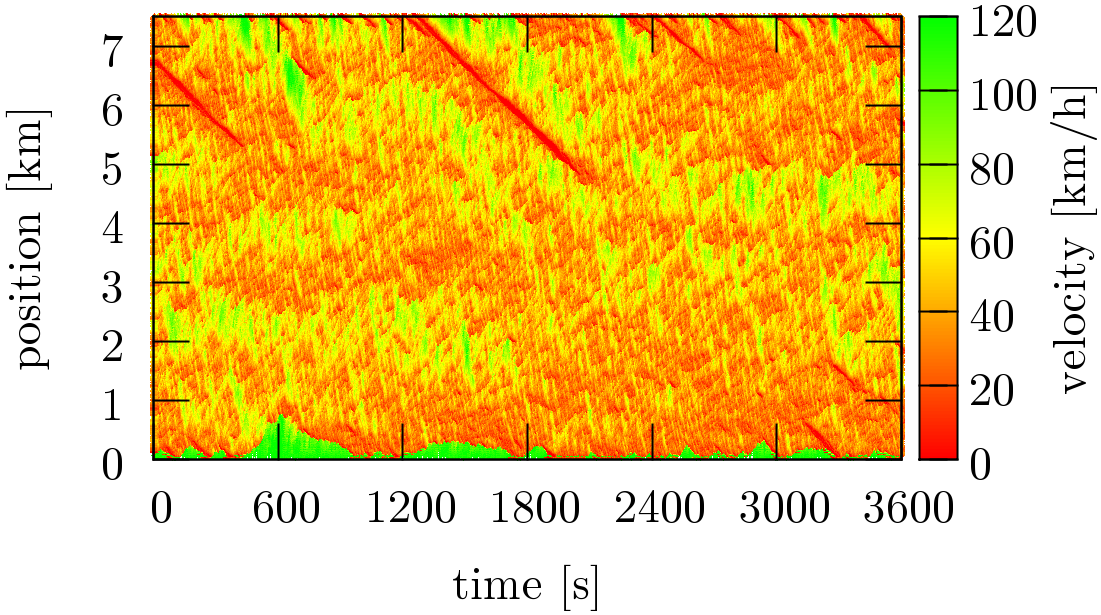}
	}\\
	\caption{Examples of different spatiotemporal patterns resulting from 
		different boundary conditions: \subref{subfig:high_flow} $\alpha = 0.86$, $\beta = 0.09$, 
									   \subref{subfig:jam_waves} $\alpha = 0.42$, $\beta = 0.32$, 
									   \subref{subfig:localized} $\alpha = 0.3$, $\beta = 0.47$, 
									   and \subref{subfig:synchronized} $\alpha = 0.38$, $\beta = 0.41$.
		(The entrance section is located at position $x=0$ and the exit section at $x=7.5$~km.)}
	\label{fig:spatiotemporal_plots} 
\end{figure}

Figure \ref{fig:spatiotemporal_plots} shows four spatiotemporal plots of the entire road during a 
one-hour interval (\ie 3600 consecutive update steps) for four distinct combinations of inflow and outflow probabilities. 
All configurations were taken from the region C in figure \ref{subfig:phase_diagram_2d}.
The pattern of figure~\ref{subfig:high_flow} is taken from the bottom right corner of figure \ref{subfig:phase_diagram_2d}. 
In this area high inflow rates coincide with a low probability of the exit section being blocked.  
The figure shows two phenomena which are characteristic for this combination of boundary conditions: 
(i) Due to the high inflow rates, random velocity fluctuations are likely to occur close to the left (upstream) boundary. 
These fluctuations cause congested traffic propagating backwards and, thereby, reduce the effective inflow probability. 
Consequently, the flow of the remaining vehicles corresponds to the outflow of congested traffic, where no other jams occur 
(see figure \ref{subfig:high_flow} during the first 1000 time steps). 
(ii) Jams, which often, but not necessarily, occur at the right boundary (at $t\approx1500$ in figure \ref{subfig:high_flow}), have a sharp 
upstream front. This is again a consequence of the high inflow rates, because any local perturbation immediately affects the 
following vehicle and propagates upstream.

Figure \ref{subfig:jam_waves} depicts several waves of congested traffic traveling upstream at a nearly constant velocity. 
These stop-and-go waves are known even from the most simplistic traffic cellular automata (\eg the Nagel-Schreckenberg model 
\cite{NagelSchreckenberg1992} or the VDR model~\cite{BarlovicSantenSchadschneiderSchreckenberg1998}).

A localized congested pattern is presented in figure~\ref{subfig:localized}. 
Relatively low exit probabilities (\ie high values of $\beta$) constantly provoke jams at the right boundary. 
The inflow probability, however, does not suffice to supply enough vehicles for the jams to propagate to the left boundary.
Hence, the jams get saturated and end close to the exit section.

\begin{figure}[hbtp]
	\subfigure[]{%
		\includegraphics[width=0.49\textwidth]{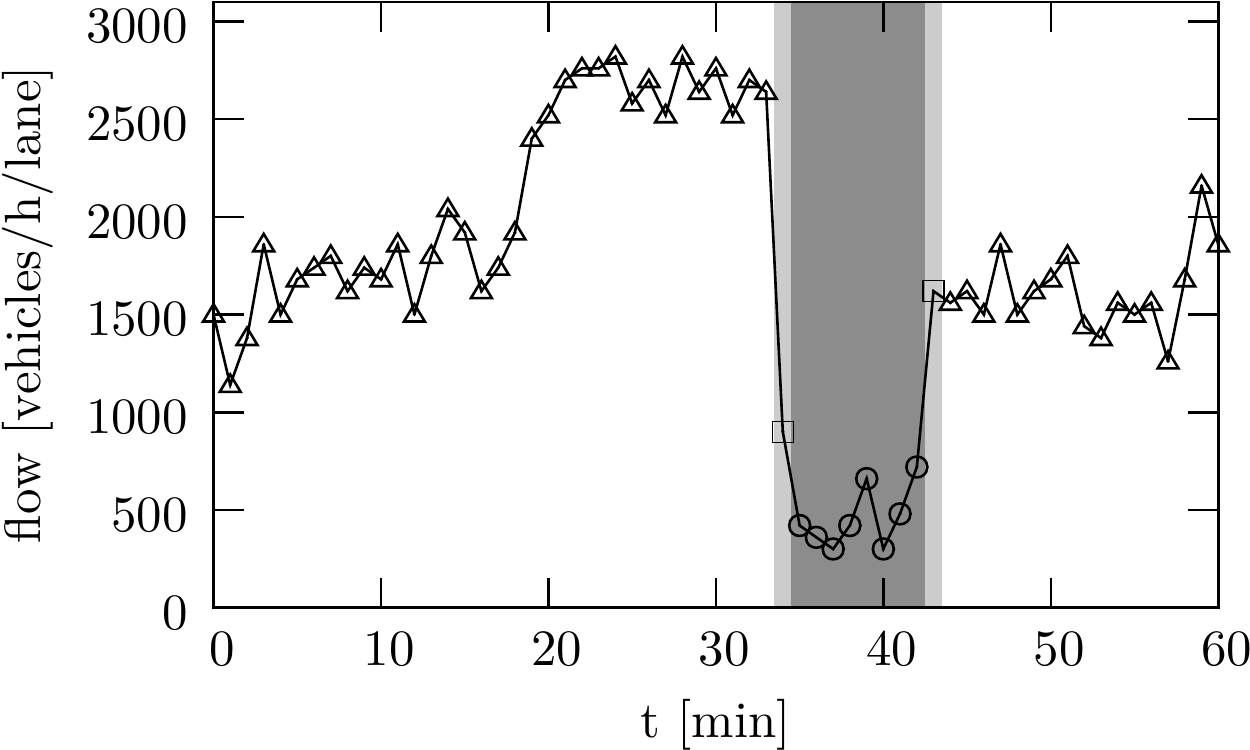}%
		\label{subfig:ts_high_flow_flow}%
	}\hfill%
	\subfigure[]{%
		\includegraphics[width=0.49\textwidth]{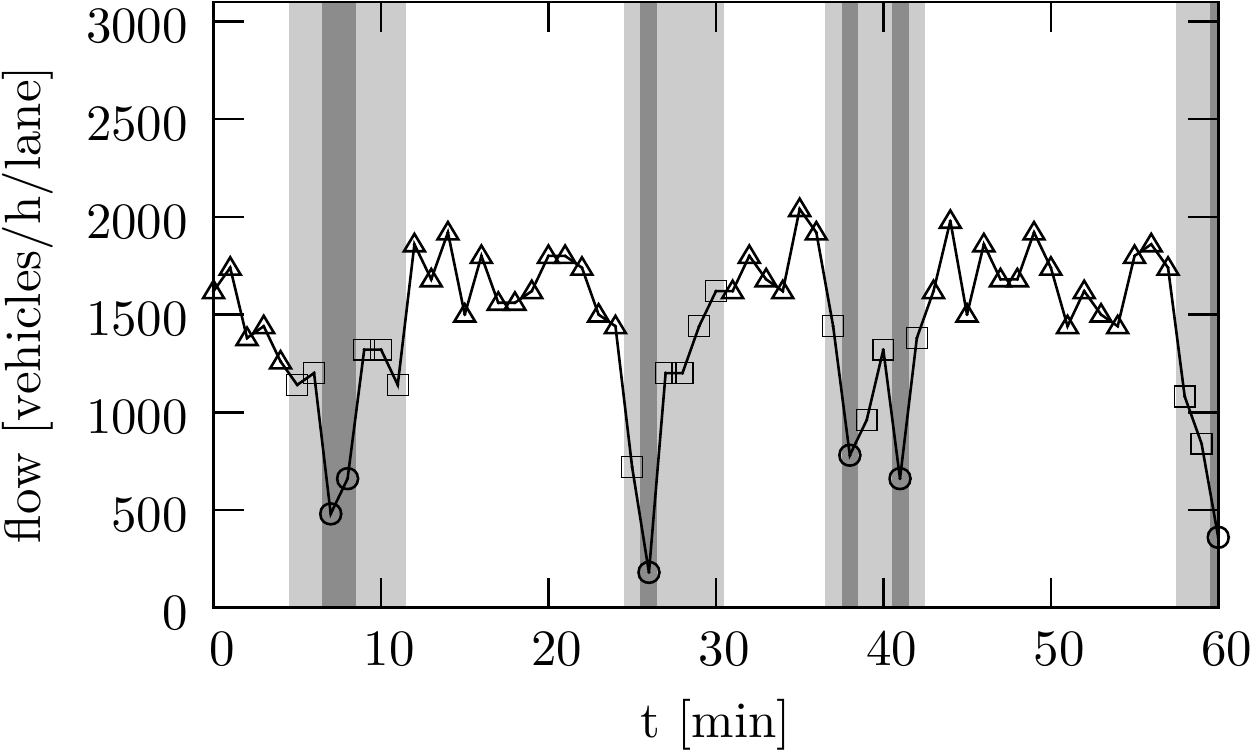}%
		\label{subfig:ts_jam_waves_flow}%
	}\\%
	\subfigure[]{%
		\includegraphics[width=0.49\textwidth]{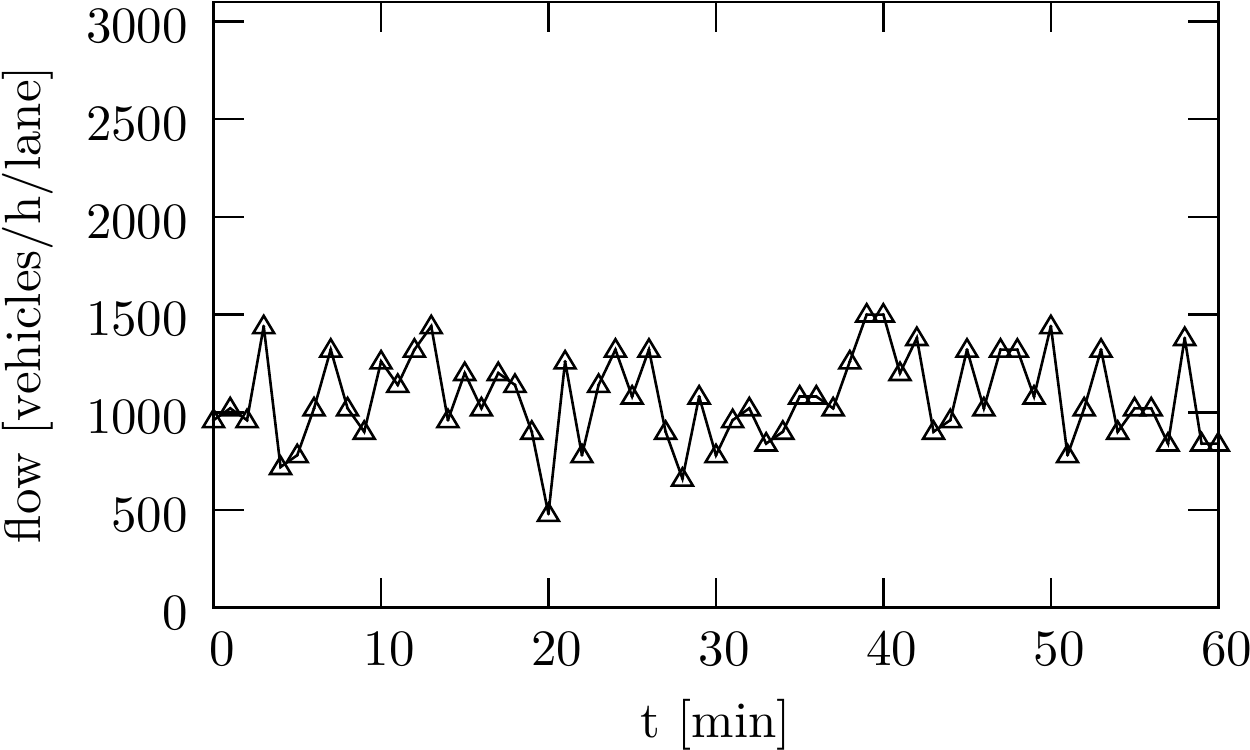}%
		\label{subfig:ts_localized_flow}%
	}\hfill%
	\subfigure[]{%
		\includegraphics[width=0.49\textwidth]{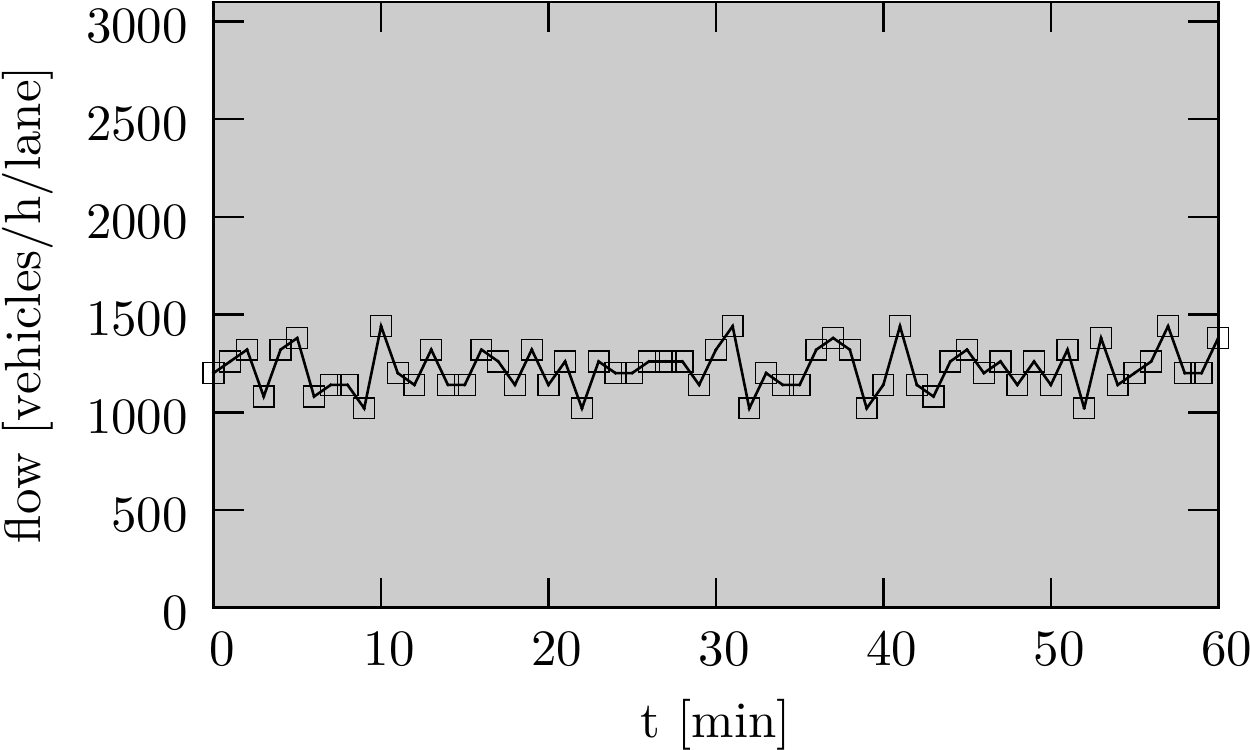}%
		\label{subfig:ts_synchronized_flow}%
	}%
	\caption{The time series of the flow rate from local measurements for the 
		patterns depicted in figure~\ref{fig:spatiotemporal_plots}. Data was collected by a detector positioned 
		in the middle of the road at kilometer $3.75$.
		Figures \subref{subfig:ts_high_flow_flow}--\subref{subfig:ts_synchronized_flow} correspond to the
		spatiotemporal plots of figure~\ref{fig:spatiotemporal_plots}:  
		\subref{subfig:ts_high_flow_flow} $\alpha = 0.86$, $\beta = 0.09$, 
		\subref{subfig:ts_jam_waves_flow} $\alpha = 0.42$, $\beta = 0.32$, 
		\subref{subfig:ts_localized_flow} $\alpha = 0.3$, $\beta = 0.47$, and 
		\subref{subfig:ts_synchronized_flow} $\alpha = 0.38$, $\beta = 0.41$.
		
		States of free flow are depicted with a white background 
		color and triangles ($\triangle$) as data points.
		Synchronized traffic is shown with a light gray background color and rectangular symbols ($\Box$).
		A dark gray background color and circular symbols ($\circ$) indicate wide moving jams. 
		
		From figures~\subref{subfig:ts_high_flow_flow} and \subref{subfig:ts_jam_waves_flow} it becomes 
		evident that there is no unique flow rate above which the traffic flow breaks down in the CDM. 
		In \subref{subfig:ts_high_flow_flow} the flow rate reaches values considerably above 
		2500\,vehicles/h/lane 
		before congestion sets in, whereas it barely exceeds 2000\,vehicles/h/lane in 
		\subref{subfig:ts_jam_waves_flow} before a breakdown occurs. 
		Also note that figures~\subref{subfig:ts_localized_flow} and \subref{subfig:ts_synchronized_flow} are assigned to 
		different traffic phases, even though the flow rates are at the same level.		
		(The corresponding velocity time series is given in figure~\ref{fig:velocity_time_series_plots}.)}
	\label{fig:flow_time_series_plots}
\end{figure}

Quite interesting is the pattern of figure~\ref{subfig:synchronized}. 
Here, nearly the entire road is covered by a state of intermediate velocities (30--70~km/h).  
At the same time, relatively high flows, ranging from 1020~vehicles/h to 1440~vehicles/h (see figure~\ref{subfig:ts_synchronized_flow}), predominate.

Such a state has not been observed for simpler models \cite{BarlovicHuisingaSchadschneiderSchreckenberg2002} 
and has led to different interpretations \cite{KernerKlenovWolf2002,JiangWu2004} for the CDM 
and a subsequent model~\cite{JiangWu2003} (see section~\ref{sec:discussion}).
The question arises: to which traffic phase(s) should one assign this spatiotemporal pattern? 
And more general: 
how can such an assignment be done for any of the above traffic patterns?

\section{Classification of Traffic Phases}
\label{sec:classification}
Kerner \etal \cite{KernerRehbornAleksicHaug2004} have presented a method called 
``FOTO'' (Forecasting of Traffic Objects) that can be used to identify traffic states (see \ref{app:foto}). 
The method uses 1-min-aggregated data from a local detector (\ie velocity and traffic flow). 
Based on a set of rules, it decides whether the local traffic state is `free flow' (F), 
`synchronized traffic' (S), or `wide moving jam' (J). 
The underlying set of rules can be summarized as follows: 
(i) if the average velocity is high, free flow predominates, (ii) if both the average velocity and the flow are low, 
a wide moving jam passes the detector, and (iii) if at medium velocities the flow is still high, then the corresponding 
traffic phase is ``synchronized flow''.
(It is important to note that the classification of traffic phases results from the simultaneous analysis of both the flow rate and 
the average velocity because the analysis of only one variable usually does not suffice to identify the traffic phase.)
In combination with a method called `ASDA' (Auto\-ma\-tische Stau-Dy\-na\-mik Ana\-lyse; Automatic Tracking of Moving Jams),
it is even possible to track the propagation of traffic phases detected by FOTO~\cite{Kerner2004} between detectors. 
An advantage of both FOTO and ASDA is that they ,,perform without any validation of model parameters in different 
environmental and traffic conditions''~\cite{KernerRehbornAleksicHaug2004}.
Hence, we could apply the FOTO-method to our simulation results without modifications.

\begin{figure}[htb]
	\subfigure[]{
		\label{subfig:ts_high_flow}
		\includegraphics[width=0.49\textwidth]{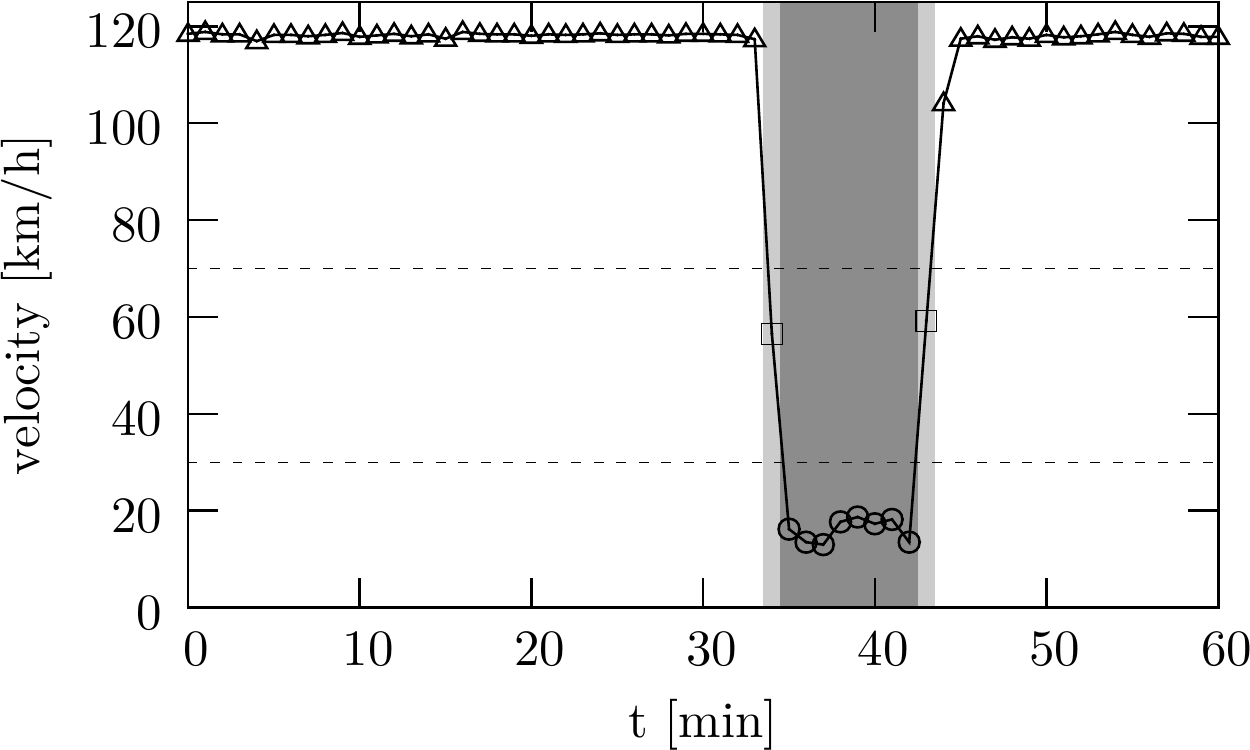}
	}\hfill
	\subfigure[]{
		\label{subfig:ts_jam_waves}
		\includegraphics[width=0.49\textwidth]{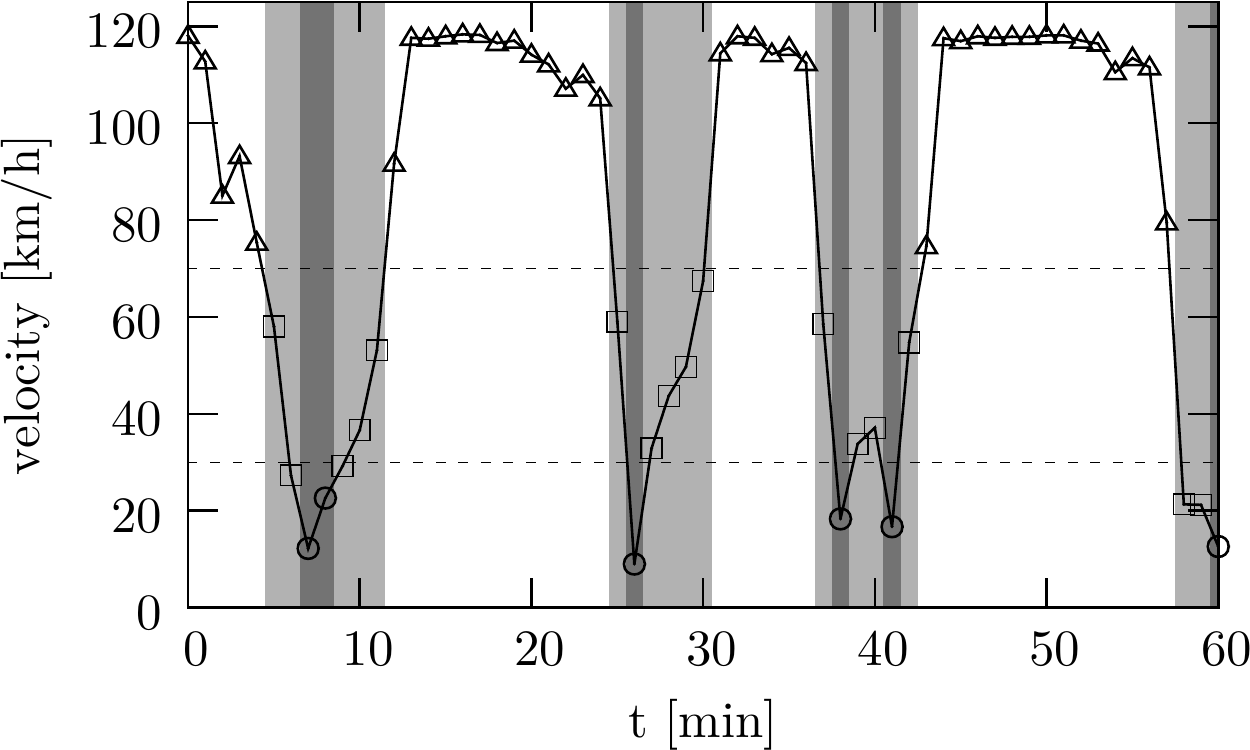}
	}\\
	\subfigure[]{
		\label{subfig:ts_localized}
		\includegraphics[width=0.49\textwidth]{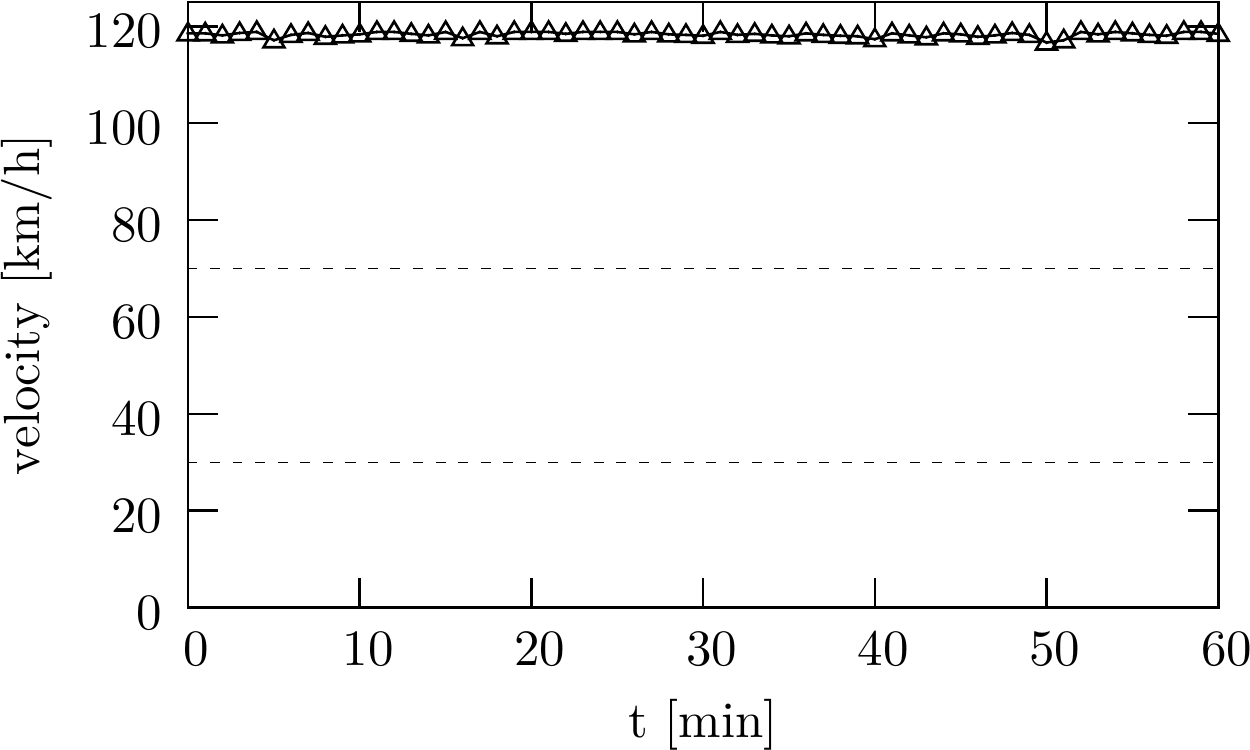}
	}\hfill
	\subfigure[]{
		\label{subfig:ts_synchronized}
		\includegraphics[width=0.49\textwidth]{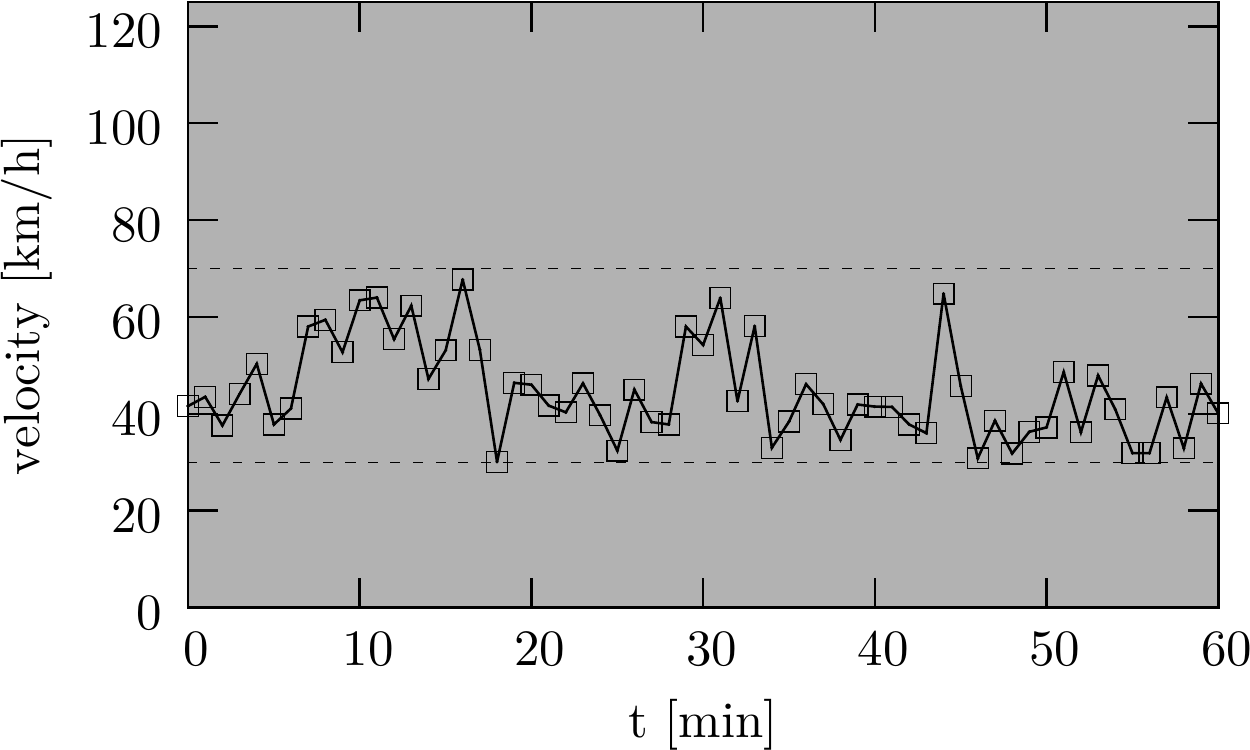}
	}\\
	\caption{The velocity time series from local measurements for 
	patterns depicted in figure \ref{fig:spatiotemporal_plots}. Data was collected by a detector positioned 
	in the middle of the road at kilometer $3.75$. (The corresponding values of $\alpha$ and $\beta$ are:
			\subref{subfig:ts_high_flow} $\alpha = 0.86$, $\beta = 0.09$, 
		\subref{subfig:ts_jam_waves} $\alpha = 0.42$, $\beta = 0.32$, 
		\subref{subfig:ts_localized} $\alpha = 0.3$, $\beta = 0.47$, and 
		\subref{subfig:ts_synchronized} $\alpha = 0.38$, $\beta = 0.41$.)
	
	Based on the local measurements of traffic flow and vehicle velocity, a classification of the current 
	traffic state was performed according to the FOTO-method. States of free flow are depicted with a white background 
	and triangles ($\triangle$) as data points.
	Synchronized traffic is shown with a light gray background color and rectangular symbols ($\Box$).
	A dark gray background color and circular symbols ($\circ$) indicate wide moving jams. 
	For better readability, dashed horizontal lines indicate the values of $30$~km/h and $70$~km/h, respectively.
	(Note that these lines are not associated with the classification of traffic states.)}	
\label{fig:velocity_time_series_plots}
\end{figure}

For this purpose, we have positioned a detector in the middle of the road (at site 2500), which gathers the same data 
as its real-world equivalent (\ie flow and velocity aggregated over 60 subsequent time steps). 
This allows a more detailed look at the vehicle dynamics. 
The velocity time series and the resulting assignment to a traffic phase for the patterns of figure~\ref{fig:spatiotemporal_plots} 
are given in figure~\ref{fig:velocity_time_series_plots}.
As expected, the drastic drop of the average velocity in figure \ref{subfig:high_flow} manifests as a wide moving 
jam, which lasts for eight minutes. Note that the detector data do not show an abrupt transition from free flow to a wide moving jam 
($\mathrm{F} \not\rightarrow \mathrm{J}$). First, we can observe a transition from free flow to a synchronized phase 
($\mathrm{F} \rightarrow \mathrm{S}$) before a wide moving jam is detected ($\mathrm{S}\rightarrow\mathrm{J}$). 
Similarly, the recovery to free flow is achieved by a sequence of two transitions 
($\mathrm{J}\rightarrow\mathrm{S}$ and $\mathrm{S}\rightarrow\mathrm{F}$).
 
The wave-like structures of figure~\ref{subfig:jam_waves} can easily be identified in figure \ref{subfig:ts_jam_waves} as well. 
All these waves show at least one aggregation interval that is identified as a jam. Again, we find the following sequence of transitions: 
$\mathrm{F} \rightarrow \mathrm{S} \rightarrow \mathrm{J} \rightarrow \mathrm{S} \rightarrow \mathrm{F}$.

As the congested states of figure \ref{subfig:localized} are located close to the exit boundary, the detector in the middle 
of the road measures only free flow (see figure~\ref{subfig:ts_localized}).

More interesting is the time series of figure \ref{subfig:ts_synchronized}, which belongs to the spatiotemporal pattern 
of figure \ref{subfig:synchronized}: 
We have already seen (figures~\ref{subfig:ts_localized_flow} and \ref{subfig:ts_synchronized_flow}) that the flow rates measured at the detector 
are approximately the same for the patterns of figures~\ref{subfig:localized} and \ref{subfig:synchronized}.
The measurements of figure~\ref{subfig:ts_synchronized} are assigned to the synchronized phase S, whereas the measurements 
of figure~\ref{subfig:localized} belong to free flow. 
The difference between the two measurements becomes evident in the velocity time series. 
In figure~\ref{subfig:ts_synchronized}, all detected velocities are between 30~km/h and 70~km/h, 
and thus they are distinctly lower than the average velocities of figure~\ref{subfig:ts_localized}.   
The maximum change during subsequent measurements is slightly below 29~km/h (from $t=43$~min to $t=44$~min).
The absolute value of these changes in the average velocity are relatively high compared with real traffic flow. 
Kerner~\cite[p.~6]{Kerner2002c}, for example, reports fluctuations in the range of $\pm 10\%$ 
at average speeds of 65~km/h.
However, the FOTO-method classified all 1-min-measurements of velocity and flow as synchronized traffic (S). 
This observation is in agreement with Kerner \cite{Kerner2011}, who attributes a self-sustaining character to synchronized traffic.
The traffic patterns at the detector's position cover the entire road, as a comparison with figure \ref{subfig:synchronized} shows. 
Consequently, we also expect the results of figure \ref{subfig:ts_synchronized} to be representative for the entire 
road---independent of the detector's position.

To get a more quantitative result on the likelihood of the transitions between traffic phases, 
we have analyzed the time series of all simulations that are not labeled `F' in figure \ref{subfig:phase_diagram_2d}. 
The exclusion of the free flow states `F' has two reasons: 
(i) In the free flow regime, where vehicles can move without hindrance, we do not expect any transitions to occur.
(ii) In very dilute traffic (\ie a small value of $\alpha$), the application of FOTO is likely to produce erroneous 
results due to very low vehicle flows. 
As very few vehicles enter the road, a detector will detect no vehicle most of the time. However,
if it does detect a vehicle, there is a high probability that no vehicle was detected during the previous sampling interval. 
This in turn, leads to a measurement of high velocity (the single vehicle travels at maximum velocity) following a measurement 
with very low velocity (no vehicle), which will be interpreted as a $\mathrm{J} \rightarrow \mathrm{F}$ transition by the FOTO-method.

In addition to the detector in the middle of the road, which we have used in the analysis of 
figure~\ref{fig:velocity_time_series_plots}, we have added another detector close to the right boundary at position $x=7.2$~km (site 4800), 
where more congestion is expected due to its proximity to the exit section.
We analyzed the time series of both detectors and classified the observed traffic states. 
The classification was performed both with the standard set of rules of FOTO, which we have 
used up to now and which comprises four distinct rules, and with an extended set of rules comprising 13 different 
rules~\cite{KernerRehbornAleksicHaug2004}, which offers a better distinction between the states J and S. 

Table~\ref{tab:transition_statistics} shows the resulting probabilities of observing a given transition. 
As we have restricted our analysis to the congested regime of figure \ref{subfig:phase_diagram_2d}, transitions from or to state J 
make up at least 70\% of all transitions at either detector location.
\begin{table}
        \caption{\label{tab:transition_statistics}Probabilities for transitions from one traffic phase
        		to another. We use both the standard and the extended rule set of FOTO. 
        		(As a consequence of rounding, the probabilities do not necessarily add up to 100\%.) }
        \begin{center}
        \begin{tabular}{lrrrr}
        	\toprule
			& \multicolumn{4}{c}{probability [\%] for the detector at}\\
         		& \multicolumn{2}{c}{$x=3.75$~km} & \multicolumn{2}{c}{$x=7.2$~km}\\
			       \cmidrule(r){2-3}\cmidrule(r){4-5}	
			transition & standard & extended & standard & extended\\
                \midrule
 				$\mathrm{J} \rightarrow \mathrm{F}$ &  0.2 &  0.2 &  0.0 &  0.3\\
				$\mathrm{J} \rightarrow \mathrm{S}$ & 44.1 & 45.3 & 35.4 & 37.6\\
 				$\mathrm{S} \rightarrow \mathrm{F}$ &  6.3 &  5.2 & 14.8 & 12.5\\
 				$\mathrm{S} \rightarrow \mathrm{J}$ & 42.8 & 44.0 & 34.9 & 36.8\\
				$\mathrm{F} \rightarrow \mathrm{S}$ &  5.0 &  3.7 & 14.4& 11.5\\
 				$\mathrm{F} \rightarrow \mathrm{J}$ &  1.6 &  1.7 &  0.4&  1.3\\
                \bottomrule
        \end{tabular}
        \end{center}
\end{table}

We found that transitions from a jammed state to free flow ($\mathrm{J} \rightarrow \mathrm{F}$) are very unlikely ($\leq 0.3\%$).
Similarly, $\mathrm{F} \rightarrow \mathrm{J}$ transitions occur with probability below 2\%.
More importantly, it has to be noted that transitions from free flow to synchronized traffic ($\mathrm{F} \rightarrow \mathrm{S}$) 
are more than two to three times more likely than $\mathrm{F} \rightarrow \mathrm{J}$ transitions. 
At the detector at position $x=7.2$~km, they are even more than eight times more likely.

This is in good agreement with empirical data. According to Kerner, spontaneous $\mathrm{F} \rightarrow \mathrm{J}$ transitions 
cannot be observed in real traffic, but wide moving jams $\mathrm{J}$ always emerge from synchronized flow~\cite{Kerner2004,Kerner2009,KernerKlenov2009}.

Concerning the different sets of rules, we can say that they led to slightly different quantitative results, but 
they did not change the qualitative character of our results.

\subsection{Analyzing Single Vehicle Data}
\label{subsec:single_vehicle_data_foto}
The findings of the previous section might lead to the conclusion that the CDM can, indeed, reproduce all the traffic phases 
proposed by Kerner.
This finding appears as a clear contradiction to the three-phase traffic theory, as the model under consideration violates the fundamental 
hypothesis of this theory. 
One might object that the usage of data aggregated over intervals of 1~min masks to some extent the inter-vehicle dynamics. 
Therefore, we felt it necessary to inspect the single vehicle data, too. 

\begin{figure}[hbtp]
	\subfigure[]{
		\label{subfig:vel_high_flow}%
		\includegraphics[width=0.44\textwidth]{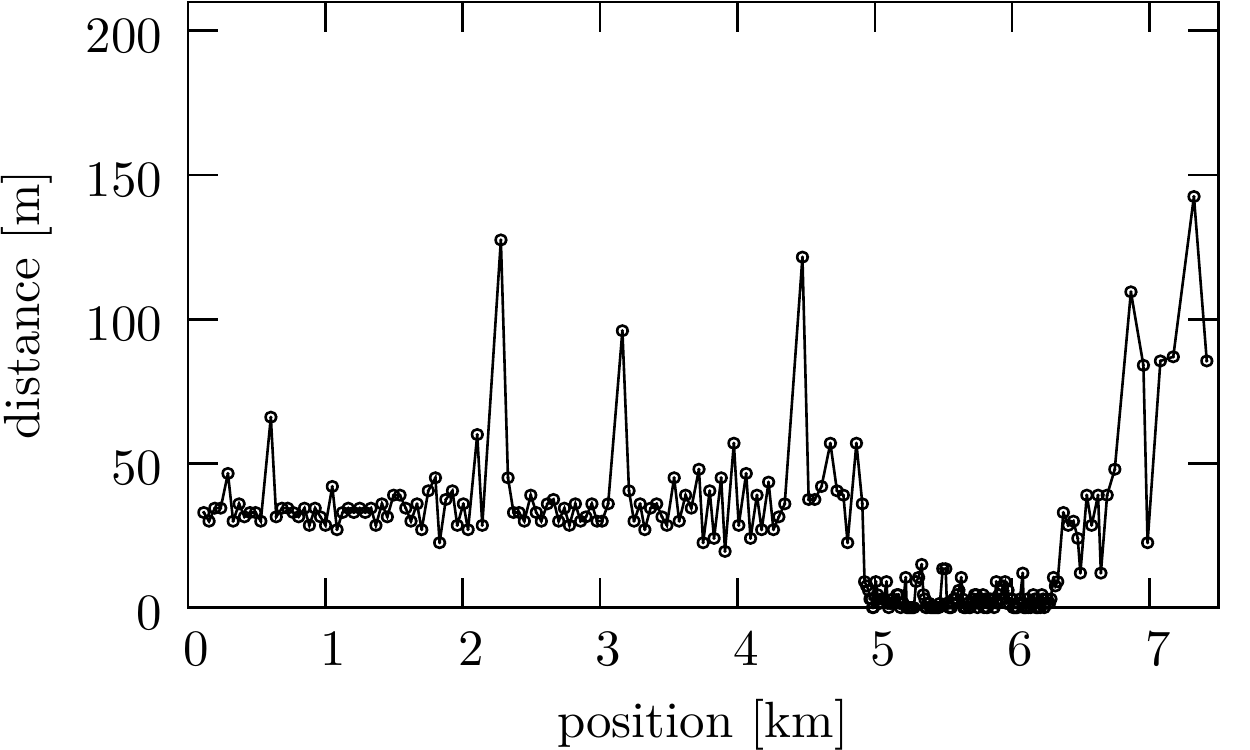}\hspace{0.08\textwidth}%
		\includegraphics[width=0.44\textwidth]{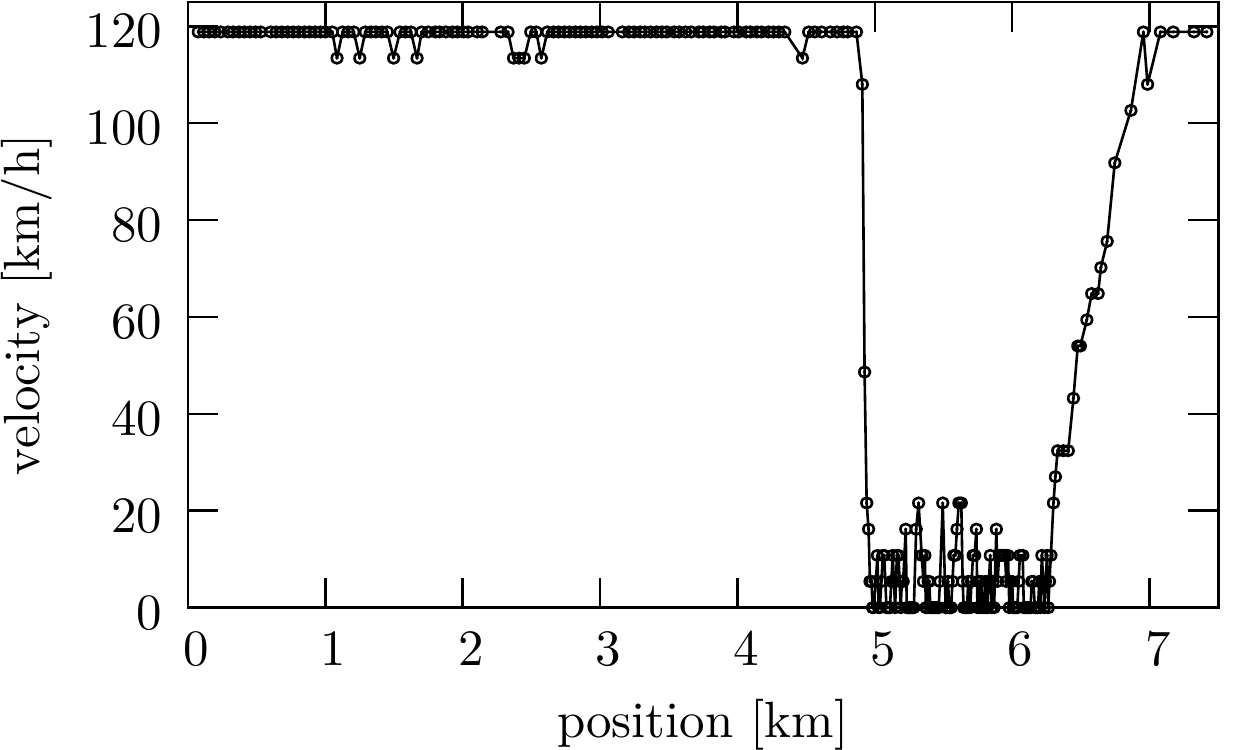}
	}\\
	\subfigure[]{
		\label{subfig:vel_jam_waves}%
		\includegraphics[width=0.44\textwidth]{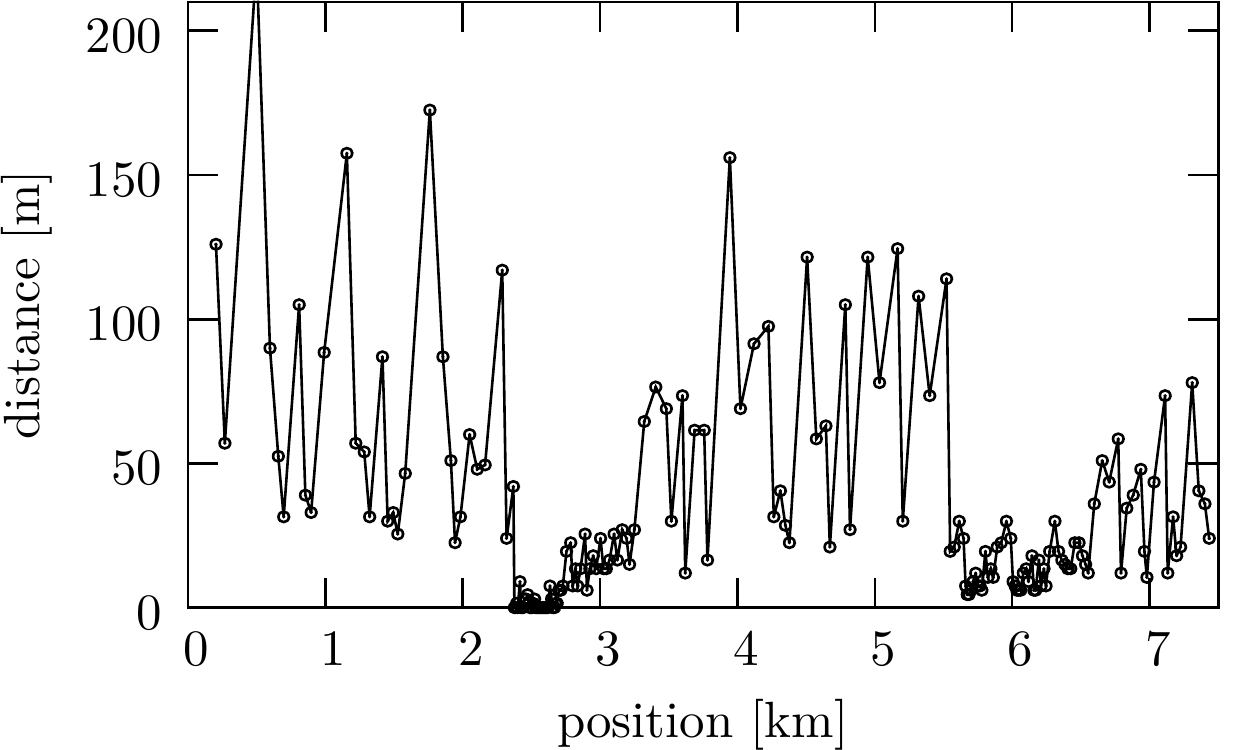}\hspace{0.08\textwidth}%
		\includegraphics[width=0.44\textwidth]{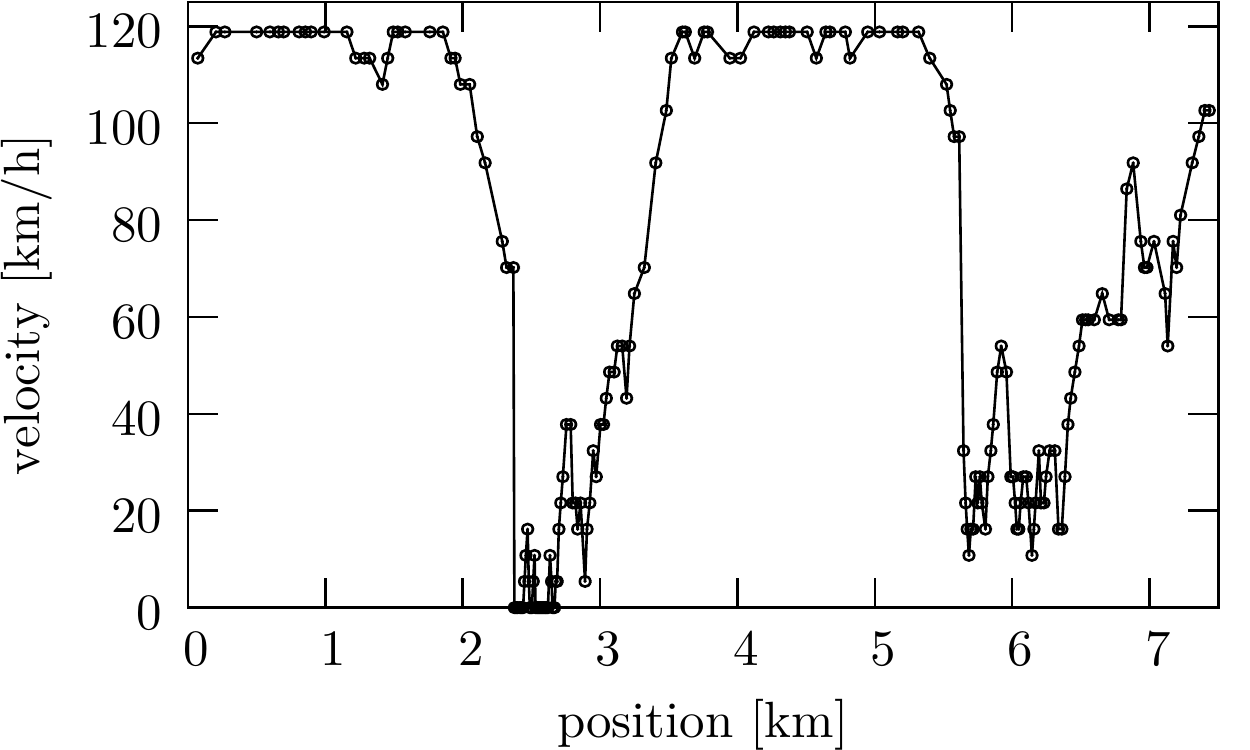}
	}\\
	\subfigure[]{
		\label{subfig:vel_localized}%
		\includegraphics[width=0.44\textwidth]{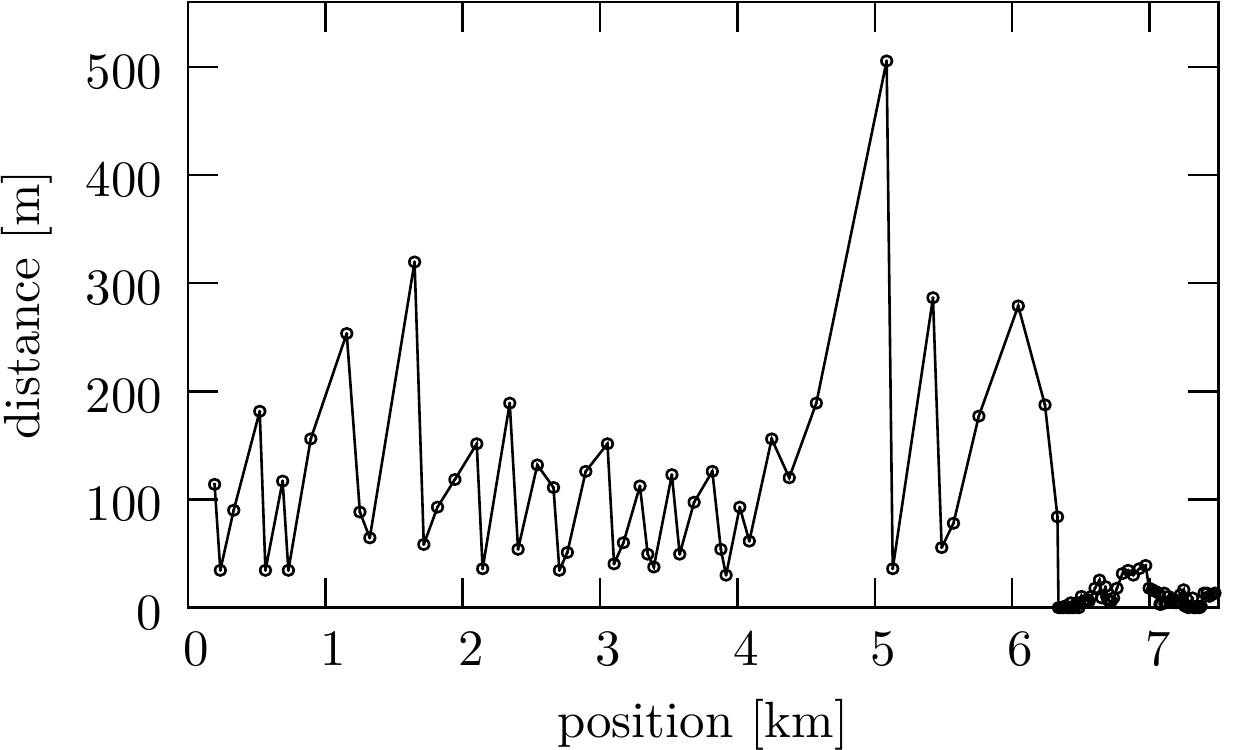}\hspace{0.08\textwidth}%
 		\includegraphics[width=0.44\textwidth]{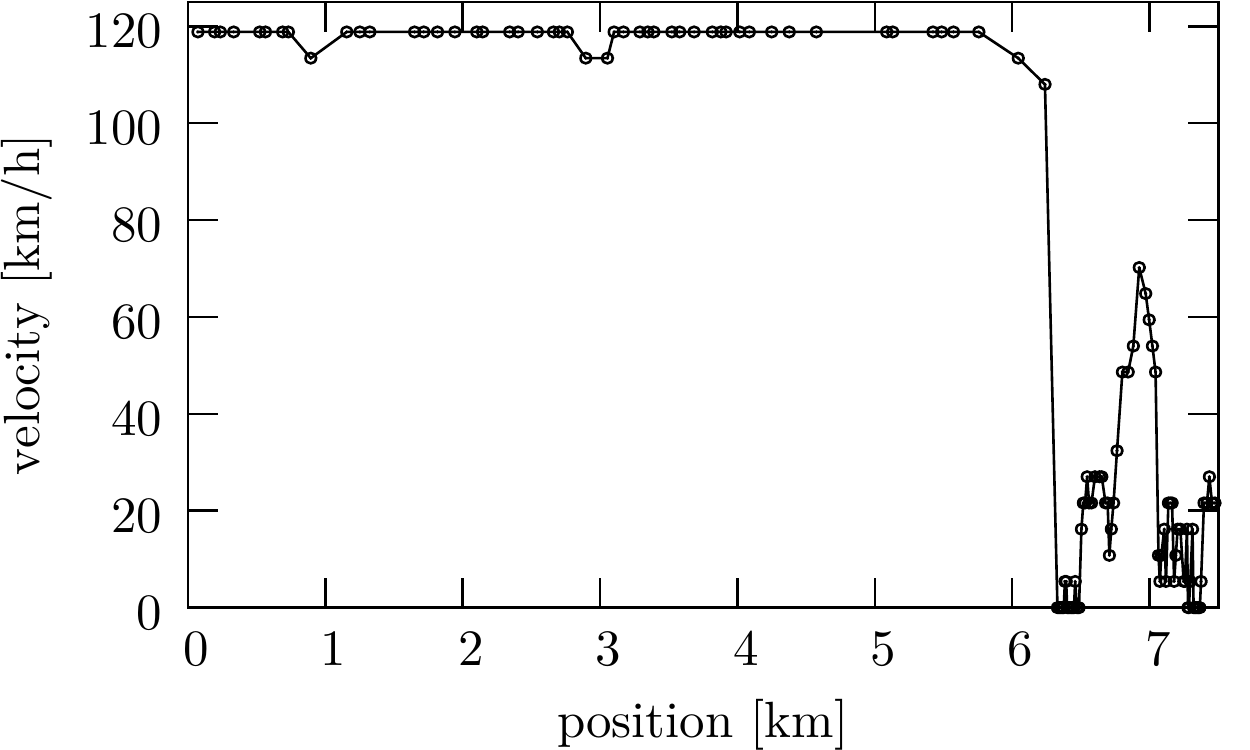}%
	}
	\subfigure[]{
		\label{subfig:vel_synchronized}%
		\includegraphics[width=0.44\textwidth]{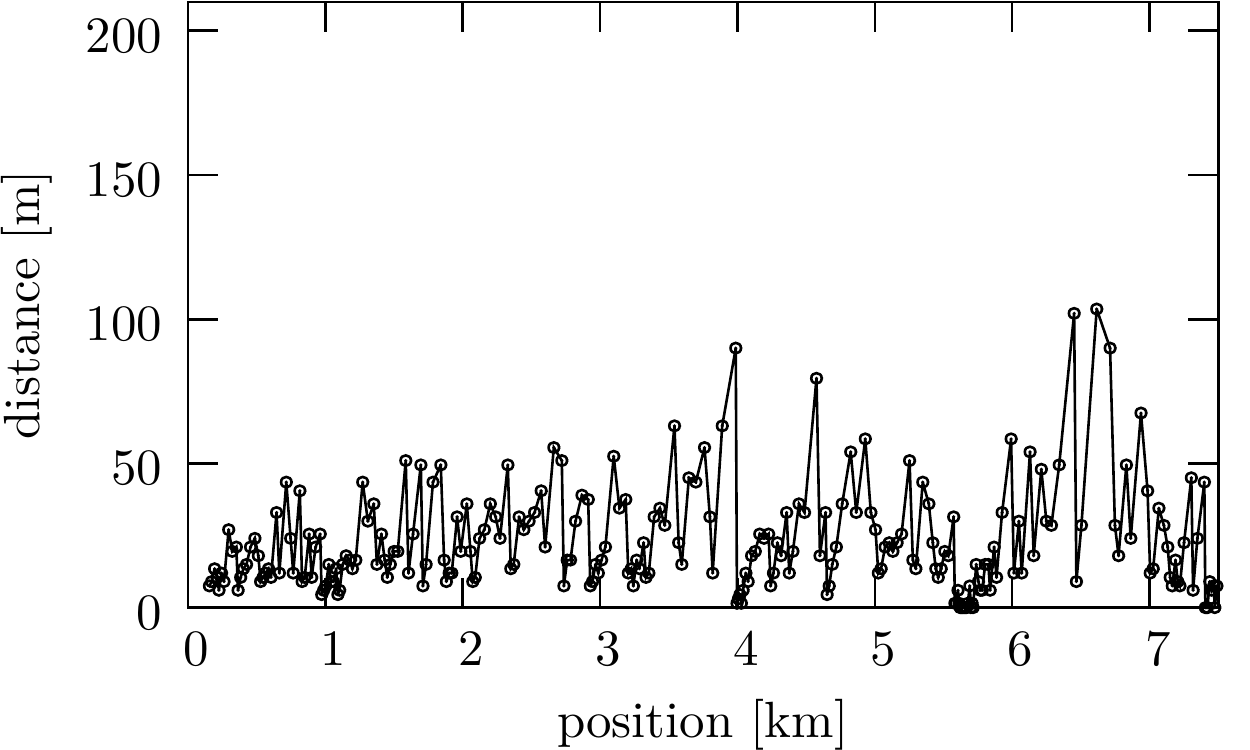}\hspace{0.08\textwidth}%
		\includegraphics[width=0.44\textwidth]{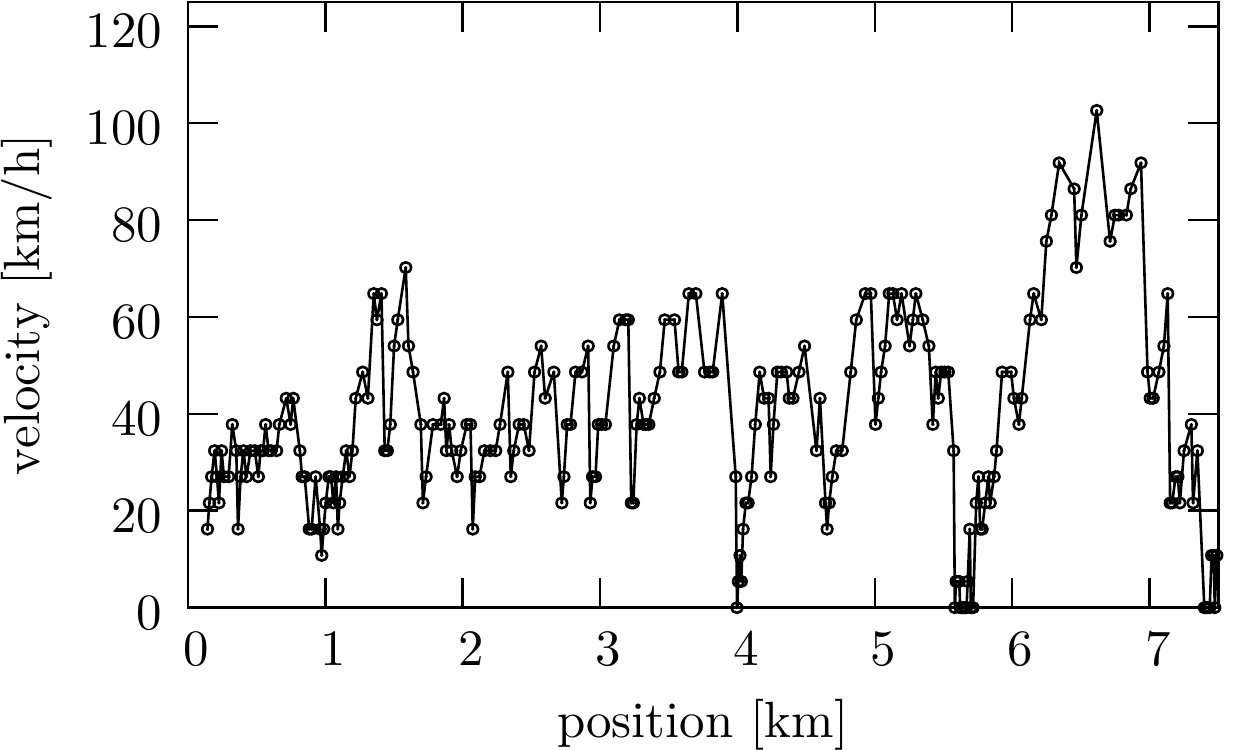}%
	}
	\caption{
	Snapshots of the vehicles' spatial headways (left) and their current velocity (right) at a fixed time $t$.
	The figures \subref{subfig:vel_high_flow}--\subref{subfig:vel_synchronized} correspond to the the simulations 
	depicted in figures~\ref{fig:spatiotemporal_plots} and \ref{fig:velocity_time_series_plots}.
	The snapshot was taken at $t=30~\textrm{min}$. 
	Note that figure~\subref{subfig:vel_localized} uses a different scale for the distance-axis. 
	(The corresponding values of $\alpha$ and $\beta$ are:
			\subref{subfig:vel_high_flow} $\alpha = 0.86$, $\beta = 0.09$, 
		\subref{subfig:vel_jam_waves} $\alpha = 0.42$, $\beta = 0.32$, 
		\subref{subfig:vel_localized} $\alpha = 0.3$, $\beta = 0.47$, and 
		\subref{subfig:vel_synchronized} $\alpha = 0.38$, $\beta = 0.41$.)}
	\label{fig:vel_ts_plots}
\end{figure}

Figure~\ref{fig:vel_ts_plots} shows a snapshot of the vehicles' headways (\ie bumper-to-bumper distance) and velocities at a fixed time $t=30~\textrm{min}$. 

As in figures~\ref{subfig:high_flow} and \ref{subfig:ts_high_flow}, the massive jam surrounded by free flow  is clearly visible in 
figure \ref{subfig:vel_high_flow}. 
Within the jam, both the vehicles' headway (\ie the bumper-to-bumper distance) and velocity are equal or close to zero. 
A similar observation can be made in figure~\ref{subfig:vel_jam_waves}, where two jam waves can be identified. 
Although, in the second jam wave, at kilometer 6, the velocity does not drop completely to zero. 
The dilute traffic of figure~\ref{subfig:vel_localized} is characterized by large headways (note the different scaling) 
and high velocities, except for the downstream boundary, where the effect of the induced perturbations becomes visible as congestion.

Figure~\ref{subfig:vel_synchronized} requires a very careful interpretation. 
The snapshot shows the headways and  velocities of vehicles in a traffic that FOTO identified as synchronized flow (figure \ref{subfig:ts_synchronized}).
We clearly see one stopped vehicle at kilometer 4 and a few (11) stopped vehicles at kilometer 5.7. 
The question of whether it is still justified to speak of the pattern of figure~\ref{subfig:vel_synchronized} as 
synchronized traffic cannot be answered definitively. 
According to the macroscopic definition of the synchronized phase, which says that any state of congested traffic that is not a 
wide moving jam is synchronized flow~\cite[p.~21]{Kerner2009}, we might call the observed traffic pattern ``synchronized''.
This implies that we do not consider 11 stopped vehicles as a wide moving jam.
On the other hand, Kerner defines synchronized traffic as traffic flow ``with no significant stoppage''~\cite[p.~23]{Kerner2009}. 
Here, the question is what constitutes a significant stoppage. 
Possible criteria are the number of stopped vehicles, the spatial extent of the stopped vehicles, or the duration of the stoppage.
As already mentioned, the number of stopped vehicles was 11, which appears to be a relatively low value compared to a wide moving jam. 
(In figure~\ref{subfig:vel_high_flow} we count 68 stopped vehicles, 
and another 34 vehicles move at 1 site per time step ($\widehat{=}\,5.4~\mathrm{km/h}$) at time $t=30\mathrm{min}$.)
Moreover, the distance between the first and last stopped vehicle in figure~\ref{subfig:vel_synchronized} was 132~m. 
Similar to the total number of stopped vehicles, we consider the spatial extent as small. 
Therefore, we also determined the time the 11 vehicles had to wait until they could move again. 
The average waiting time was 14~s, and the longest waiting time was 25~s. 

Studying the vehicles' time headways allows for a more objective analysis: 
Kerner \etal\cite{KernerKlenovHillerRehborn2006} reported that one finds regions of interrupted flow within wide moving jams. 
These flow interruptions are characterized by maximum time headways of $t_\mathrm{h,max}\geq20$~s between two vehicles.  
So, for each stopped vehicle, we recorded the time headways that a local detector at the corresponding vehicle's position would have measured.
The average time headway of the next ten following vehicles was 4~s. Yet, once, a time headway of 33~s was observed. 
All other time headways were below 8~s.
Consequently, based on Kerner's microscopic criterion $t_\mathrm{h,max}\geq20~s$, 
we would have to reject the classification of the FOTO-method for some time intervals. 
Moreover, we would have to conclude that the pattern of figure~\ref{subfig:synchronized} 
is not entirely jam-free, even if no wide moving jam could be identified via FOTO. 

At this point, it has to be mentioned that the three-phase traffic theory also treats so-called narrow moving jams~\cite[p.~259]{Kerner2009}, which, in 
contrast to wide moving jams, consist merely of an upstream and downstream front. 
These narrow moving jams can either grow into wide moving jams or disappear completely, and they are associated with the synchronized flow phase. 
Hence, if the aforementioned sequence of stopped vehicles represents a narrow moving jam, which its 
length of 132~m suggests, then the FOTO-method's classification as synchronized flow is still correct.

Considering all the previous points, a final answer on how to interpret the traffic pattern 
of figure~\ref{subfig:synchronized} is not possible. 
On the one hand, one might argue that the single large time-headway was a statistical fluctuation rather 
than a proof of the existence of a wide moving jam. 
On the other hand, one might object that only single vehicle data are a reliable source for the analysis 
of traffic patterns because detector data provide incomplete information due to the aggregation process.
Yet the sequence of stopped vehicles covers less than 2\% of the considered road segment.
Therefore, the identification of the observed stoppage as a wide moving jam appears not mandatory, and 
its identification as a narrow moving jam seems justified as well.

\subsection{Simulated versus Empirical Data}
\label{subsec:simulated_vs_empirical}
\begin{figure}[hbtp]
	\centering
	\subfigure[]{%
		\label{subfig:highway}%
		\includegraphics[width=1\textwidth]{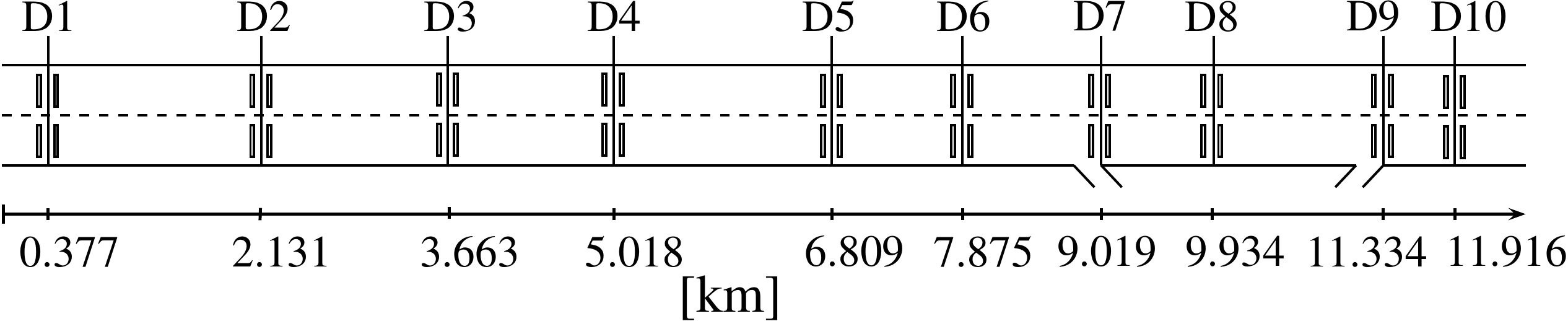}%
	}\\%
	\subfigure[]{%
		\label{subfig:breakdown}%
		\includegraphics[width=1\textwidth]{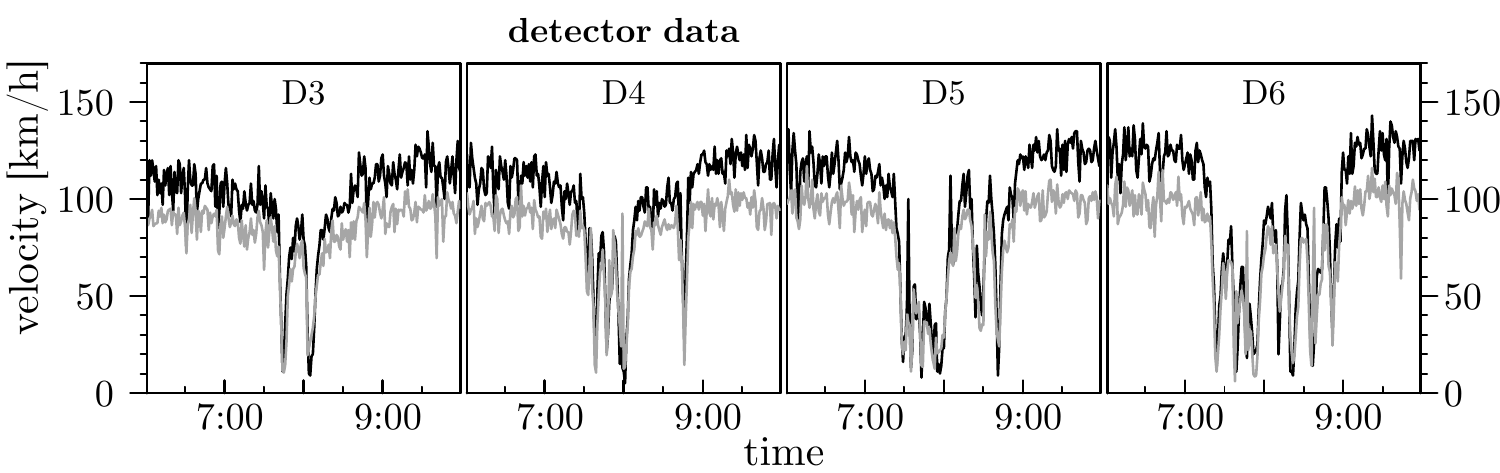}%
	}%
	\caption
	{
	Empirical data used for a comparison with simulated data were taken from several detectors on German Autobahn A44. 
	The Autobahn-section considered is depicted in (a), where the detector cross-sections are labeled D1,\ldots,D10.
	Time series of the average velocity for detectors D3--D6 are shown in (b). 
	A solid black line represents the time series of cars, whereas the time series of trucks is shown with a solid gray line.
	}
	\label{fig:highway_and_breakdown}
\end{figure}

In contrast to the simplistic setup of section~\ref{sec:open_boundaries}, we finally want to provide some 
more realistic results.
Therefore, we apply the FOTO-method to both empirical and simulated traffic data. 
In a recent article~\cite{KnorrSchreckenberg2012c}, the authors investigated the ability of three 
microscopic traffic models---including the CDM---to reproduce a traffic breakdown, \ie the abrupt and 
spontaneous transition from free flow to congested traffic. 
This investigation was based on empirical data collected by 10 detectors on the German Autobahn A44, as 
sketched in figure~\ref{subfig:highway}. 
In great detail we have simulated the depicted two-lane Autobahn-segment with two types of vehicles (\ie cars and trucks). 
The flow rates for both cars and trucks follow the rates measured by the detector D1.  
From the detectors' time series of November 4, 2010, shown in figure~\ref{subfig:breakdown}, 
this breakdown can easily be identified in the  morning peak hour between 7~a.m.\ and 9~a.m.

In the following, we will contrast these empirical measurements with data obtained from simulations 
with the CDM. 
The computer simulations emulated the highway segment of figure~\ref{subfig:highway}, and the inflow and outflow via 
the boundaries (including the ramps) followed the empirical data. 
(For a detailed description of the simulation setup, see~\cite{KnorrSchreckenberg2012c}.)

To apply the FOTO-method, we have averaged the vehicle flow across lanes, and we have calculated the average 
velocity by a weighted average of the velocity of trucks and vehicles over both lanes. 

\begin{figure}[hbtp]
	\subfigure[]{
		\label{subfig:foto_ts_5018_join1}%
		\includegraphics[width=0.46\textwidth]{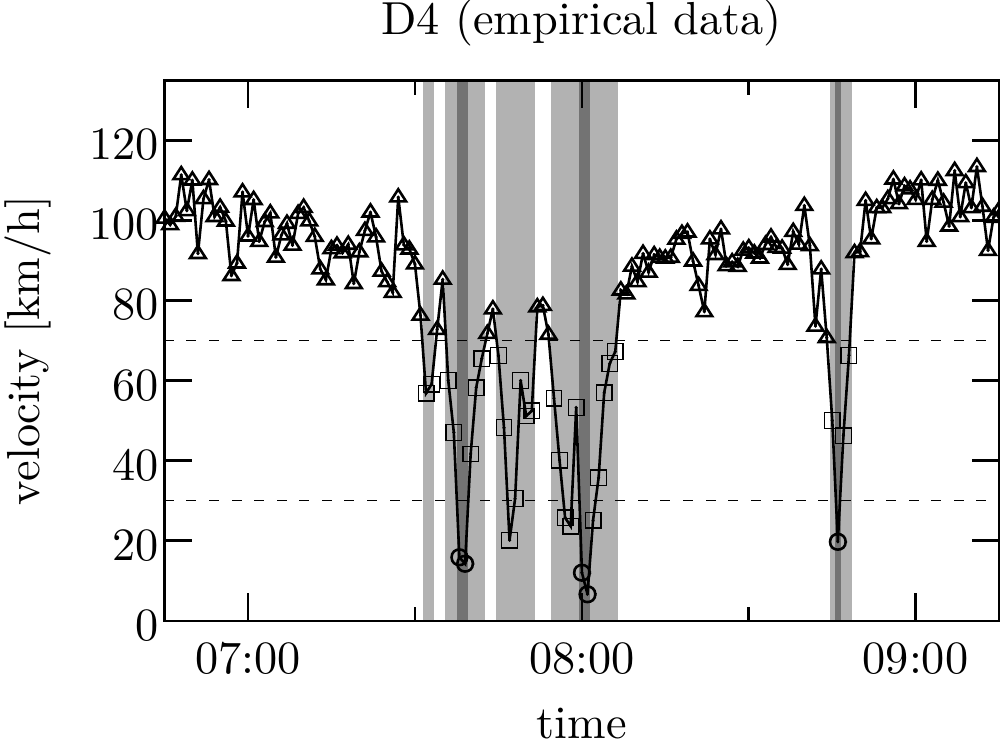}\hspace{0.06\textwidth}%
		\includegraphics[width=0.46\textwidth]{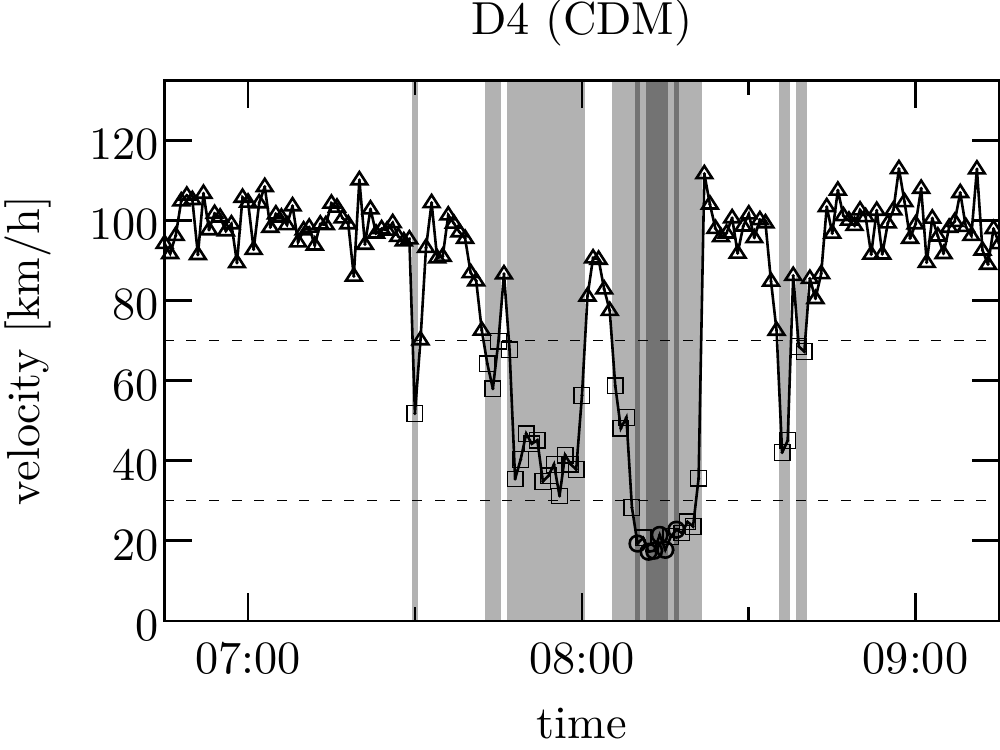}%
	}\\%
	\subfigure[]{
		\label{subfig:foto_ts_6809_join1}%
		\includegraphics[width=0.46\textwidth]{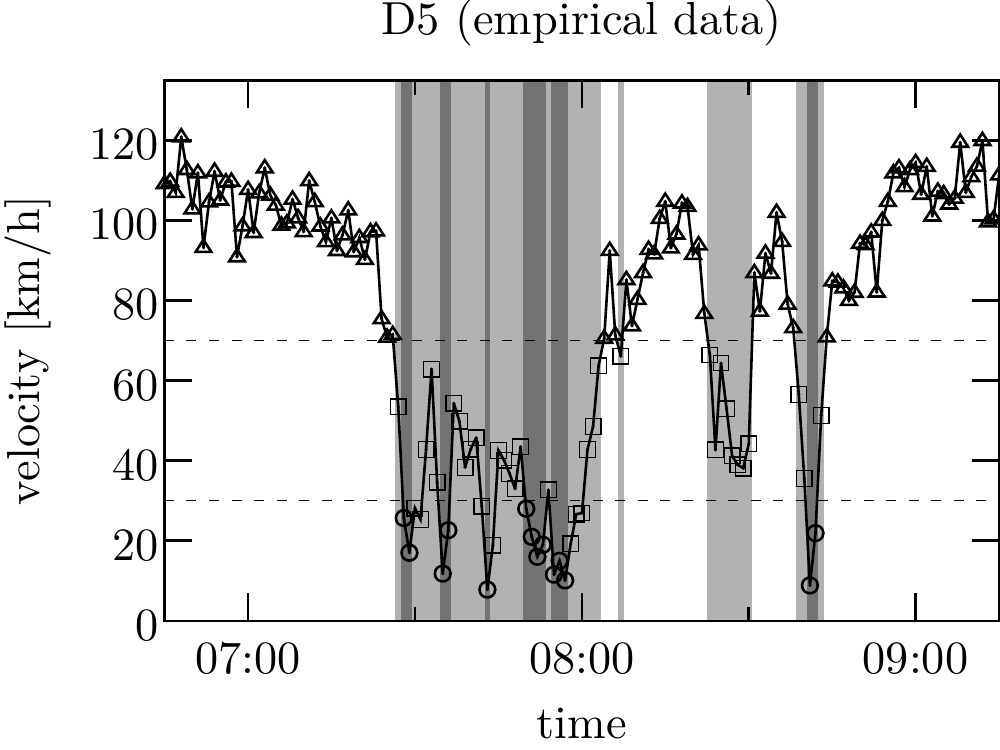}\hspace{0.06\textwidth}%
		\includegraphics[width=0.46\textwidth]{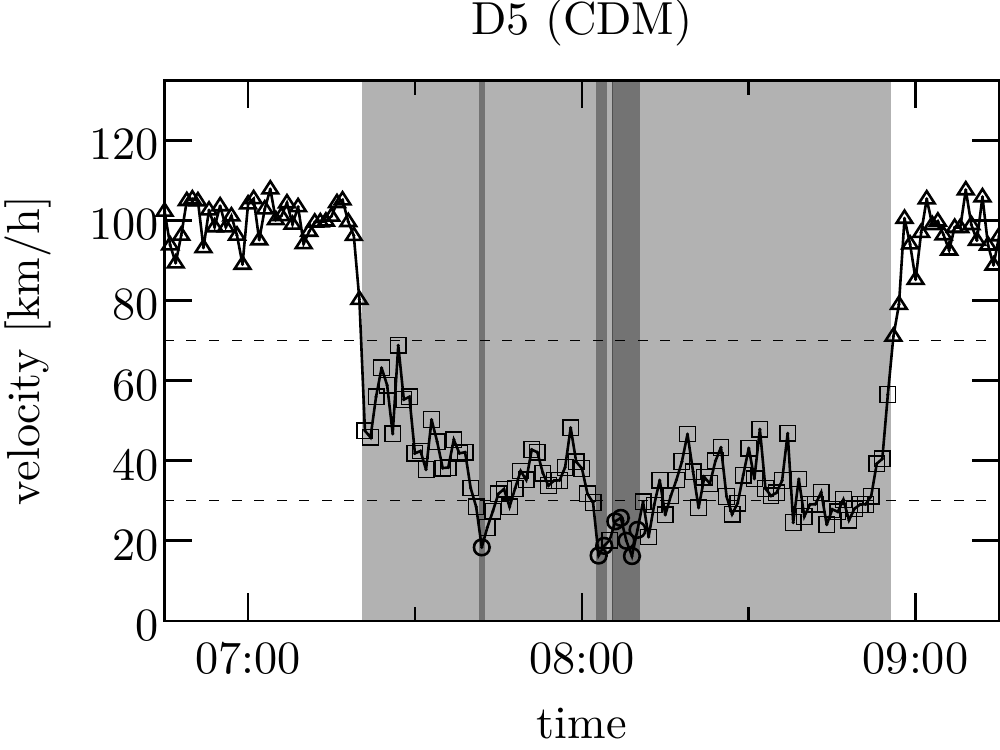}
	}\\%
	\subfigure[]{
		\label{subfig:foto_ts_7875_join1}%
		\includegraphics[width=0.46\textwidth]{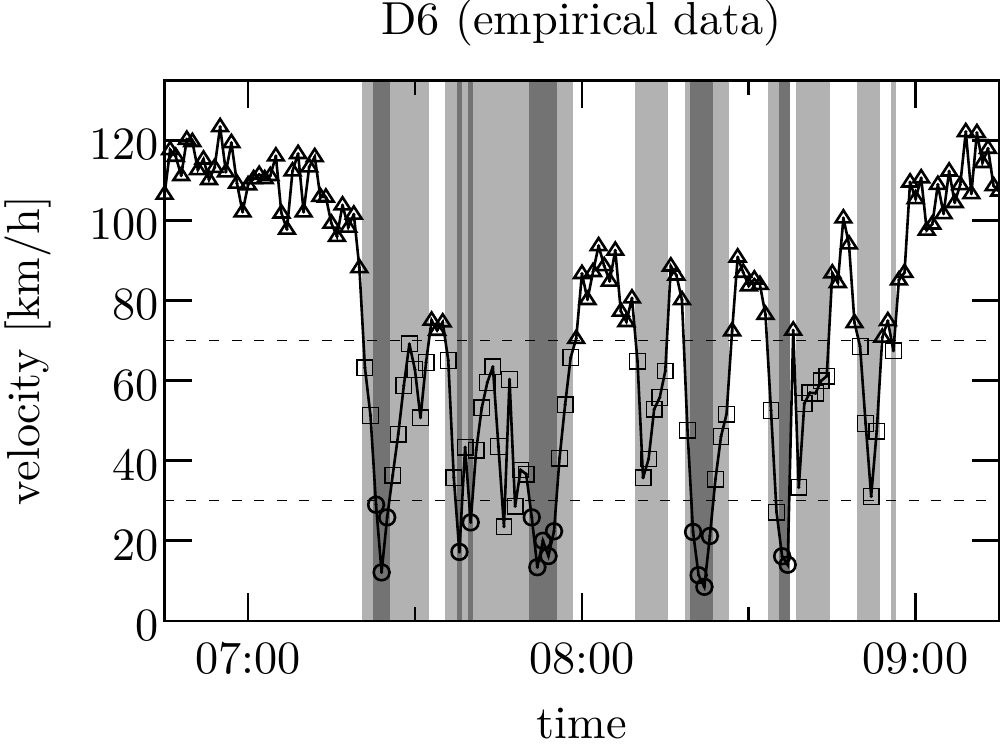}\hspace{0.06\textwidth}%
		\includegraphics[width=0.46\textwidth]{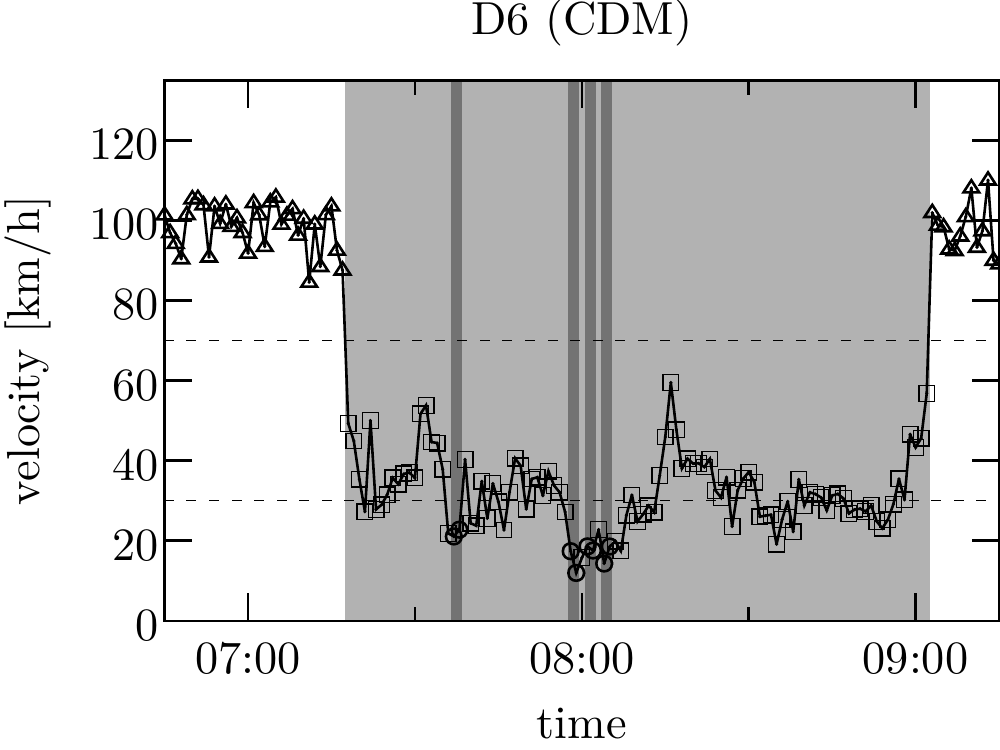}
	}%
	\caption{
	A comparison of the traffic state classification according to the FOTO-method for both the empirical (left) 
	and the simulated (right) time series.
	
	States of free flow are depicted with a white background color and triangles ($\triangle$) as data points.
	Synchronized traffic is shown with a light gray background and rectangular symbols ($\Box$).
	A dark gray background color and circular symbols ($\circ$) indicate wide moving jams. 
	}
	\label{fig:foto_empirical_vs_cdm}
\end{figure}

The resulting classification of traffic states is given in figures~\ref{subfig:foto_ts_5018_join1}, \ref{subfig:foto_ts_6809_join1}, 
and \ref{subfig:foto_ts_7875_join1} for the detector cross-sections D4, D5, and D6, respectively (see figure~\ref{subfig:highway}). 
The plots on the left show the empirical data, and the plots on the right show the detectors' time series from simulations with the CDM.

The comparison between real and simulated data confirms some earlier findings~\cite{KnorrSchreckenberg2012c}: 
the CDM overestimates the temporal extent of congested traffic during the morning peak hour. 
At detectors D5 and D6, an uninterrupted sequence of congested traffic could be found between 7:20~a.m.\ and 8:55~a.m.\ for  the CDM.  
The real time series at these locations show free flow during a 10~min-interval at approximately 8:10~a.m. 

Table~\ref{table:classification_empirical_cmd}, which lists the number of intervals assigned to the traffic phase J, F, and S, 
confirms this observation as well, but allows for a quantitative characterization. 
On the one hand, the simulation results overestimate the occurrence of congested traffic, but they underestimate the 
number of intervals with wide moving jams (J). 
\begin{table}[hbtp]
	\caption{
	Overview of the traffic phases found in the time from 6:00~a.m.\ to 10:00~a.m.\ at the detectors downstream of 
	the off-ramp.}
	\label{table:classification_empirical_cmd}
	\centering
	\begin{tabular}{lrrrrrr}
	\toprule
		  & \multicolumn{6}{c}{1~min detector data}\\
		  \cmidrule(r){2-7}
    	  & \multicolumn{3}{c}{empirical} & \multicolumn{3}{c}{simulated}\\
	      \cmidrule(r){2-4}\cmidrule(r){5-7}
	detector & {J} & {S} & {F} & {J} & {S} & {F}\\
	\midrule
	D6 & 16 & 44 & 177 & 8 & 97 & 136 \\
	D5 & 14 & 37 & 190 & 8 & 87 & 146 \\
	D4 &  5 & 27 & 209 & 6 & 32 & 203  \\
	D3 &  9 & 13 & 219 & 0 &  7 & 234  \\
	\bottomrule
\end{tabular}
\end{table}

The comparison between the empirical and simulated time series of figure~\ref{fig:foto_empirical_vs_cdm} 
as well as with the time series of figure~\ref{fig:velocity_time_series_plots} is revealing for another reason: 
in figures~\ref{subfig:ts_high_flow} and \ref{subfig:ts_jam_waves} the transition F$\rightarrow$S and 
the subsequent S$\rightarrow$J transition were only one or two minutes apart. 
Hence, one might assume that the intermediate step S is a mere artifact of the aggregation process. 
In figures~\ref{subfig:foto_ts_6809_join1} and \ref{subfig:foto_ts_7875_join1}, however, the state 
classified as S by the FOTO-method lasts for more than 10~min before the S$\rightarrow$J follows. 
On the other hand, we observe such short-lasting S-states in the time series of the real detectors in 
figures~\ref{subfig:foto_ts_5018_join1}--\ref{subfig:foto_ts_7875_join1}. 
This proves that the CDM is also able to exhibit long-lasting intervals of synchronized flow that 
precede a wide moving jam.

It should be noted that a few F$\rightarrow$J transitions could be found  in the empirical data 
as well. 
To ensure that this effect did not result from the chosen averaging process, we also applied 
the FOTO-method to the raw data. 
The analysis of the raw data confirmed the existence of F$\rightarrow$J transitions.
This observation, which contradicts three-phase theory, is rather an evidence 
of an imperfection of the FOTO-method than of the theory itself. 
Remember that the traffic phases are spatiotemporal phenomena, whereas the FOTO-method 
relies on local information only. 

\section{Discussion}
\label{sec:discussion}
In this article, we studied the spatiotemporal dynamics of the comfortable driving model (CDM). 
We felt such an analysis was necessary, as many newly presented traffic models (\eg \cite{JiangWu2004,ShangPeng2012}) claim 
to reproduce the synchronized phase of Kerner's three-phase traffic theory simply by 
presenting space-time plots such as the ones shown in figure~\ref{fig:spatiotemporal_plots}. 
Hence, we used the rule-based FOTO-method which, although it certainly has some limitations in assessing the 
spatiotemporal dynamics as it is based on local measurements only, still provides hard and objective criteria for this purpose. 

In the article by Kerner \etal \cite{KernerKlenovWolf2002}, for instance, in which also the CDM was investigated, 
the spatiotemporal pattern of figure \ref{subfig:synchronized} was 
classified as an `oscillating moving jam'. It appears, however, that the same pattern was classified as synchronized flow 
by Jiang and Wu \cite[Figs. 3+4]{JiangWu2004}, who analyzed an extension of the CDM \cite{JiangWu2003}.
(Their extended model incorporates some findings of the three-phase traffic theory.)
This observation only confirms the difficulty of classifying traffic phases which we have mentioned in the
introductory section. 
Therefore, a ruled-based method such as the FOTO-method is preferable, 
as it promises an objective classification and facilitates the comparability of our results with  
both empirical data (figure~\ref{fig:foto_empirical_vs_cdm}) and other traffic models.

By applying the FOTO-method to the CDM, we found that the obtained results are in both good quantitative and 
qualitative agreement with findings of the three-phase traffic theory. 
In particular, we have demonstrated that the CDM exhibits three clearly distinguishable traffic states or ``phases''.
Notwithstanding these results, it is still not clear whether the CDM 
reproduces the synchronized phase in the sense of Kerner's three-phase traffic theory. 
For, according to Kerner, any traffic model with a functional relationship between traffic flow and 
density, such as the CDM, cannot adequately describe synchronized flow \cite{KernerKlenovWolf2002}, 
where such a relationship does not exist. 
At this point, it has to be mentioned that traffic states where traffic density and flow are practically uncorrelated exist in the 
CDM as well~\cite{KnospeSantenSchadschneiderSchreckenberg2004}. 
This uncorrelated behavior is a consequence of strong fluctuations of flow, velocity, and density, and does 
not contradict the existence of a flow-density relationship.

Despite the overall good agreement with empirical data, we want to point out a 
major difference of the CDM and models within the three-phase theory:
In contrast to the latter, the fluctuations of velocity (figure \ref{subfig:ts_synchronized}) 
are considerably higher for the CDM (\eg cf. \cite[Fig. 9]{KernerKlenovWolf2002}). 
This is a consequence of the different modeling approaches. In early traffic cellular 
automata (\eg \cite{NagelSchreckenberg1992,BarlovicHuisingaSchadschneiderSchreckenberg2002}), a vehicle always 
accelerates when it is safe to do so. In the CDM this effect is partially compensated by an anticipatory component. 
In models based on three-phase traffic theory (\eg the models presented in \cite{KernerKlenovWolf2002}), a vehicle 
may accept any gap if its velocity is within a given range that is determined by the models' rules of motion.

Finally, let us emphasize that we fully acknowledge fundamental differences between models based on three-phase 
traffic theory and the fundamental diagram approach: The CDM, for example, fails to explain the microscopic 
origin of synchronized traffic (\ie the microscopic mechanisms leading to a F$\rightarrow$S transition as found by Kerner~\cite[ch.~3]{Kerner2009}), 
although it correctly reproduces many aspects of the three-phase theory. 
Therefore, its application to large scale traffic networks is still justified \cite{Olsim}.
\section*{Acknowledgments}
FK thanks A. Schadschneider for fruitful discussions and the German Research Foundation (DFG) 
for funding under grant number SCHR 527/5-1. 
Both authors thank the anonymous reviewers for their helpful comments.

\begin{appendix}
\section{The FOTO-method}
\label{app:foto}
\begin{figure}[hbtp]
\centering
 \subfigure[]{
   \includegraphics[width=0.49\textwidth]{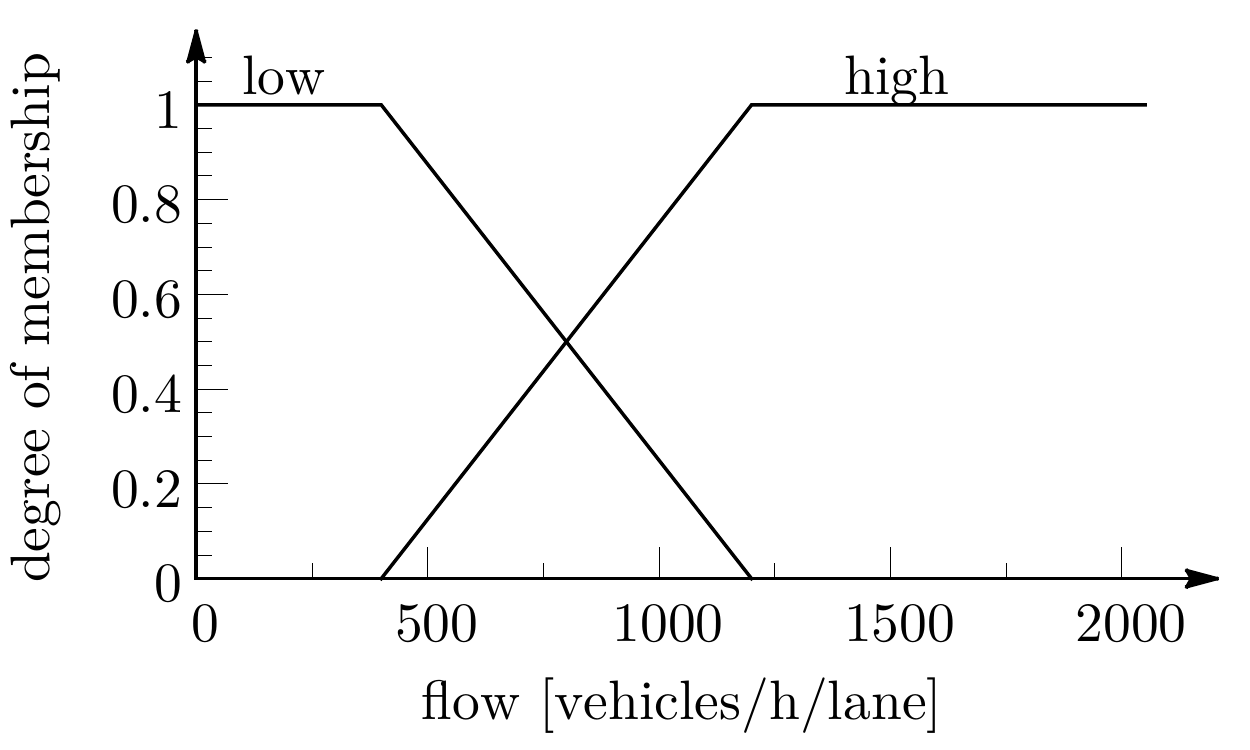}
   \label{fig:foto_flow}
 }%
 \subfigure[]{
   \includegraphics[width=0.49\textwidth]{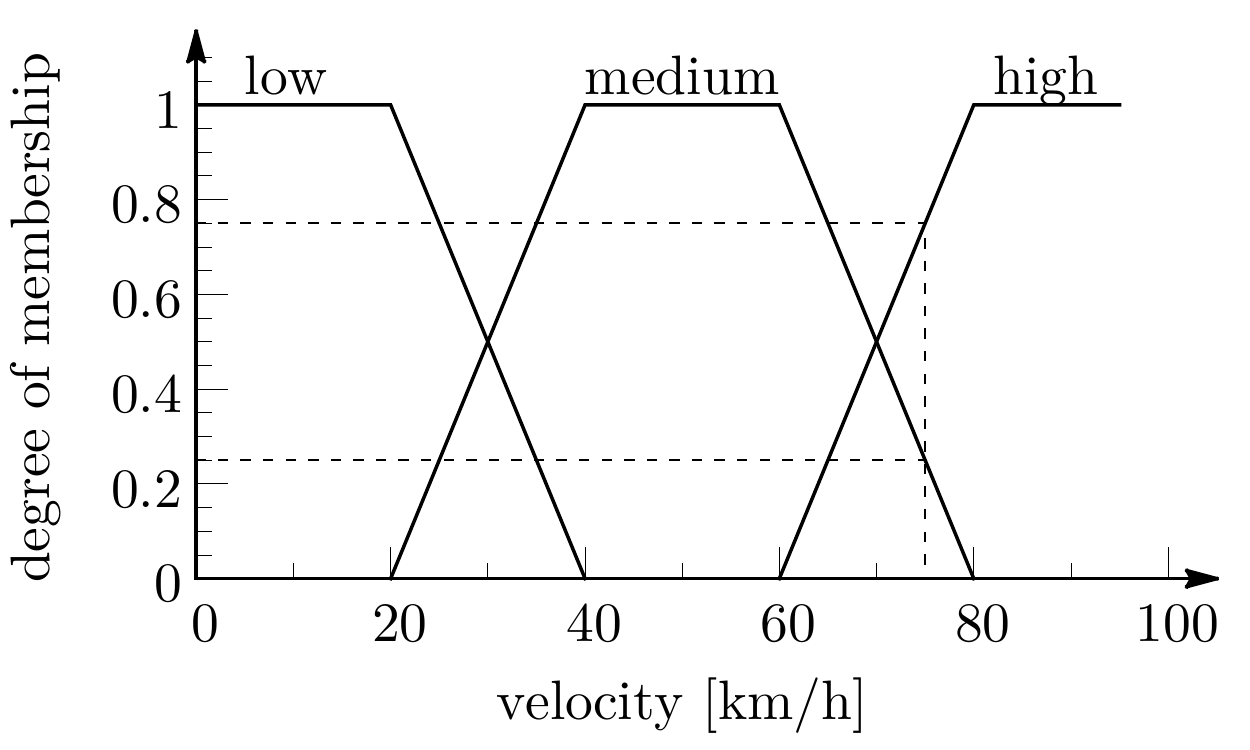}
   \label{fig:foto_velocity}
 }%
\caption
	{Illustration of the fuzzification process. 
	For each value of \subref{fig:foto_flow} flow and \subref{fig:foto_velocity} velocity one can read off the 
	associated degree of membership to the classes ``low'', ``high'', and, in the case of the velocity, ``medium''. 
	The dashed lines in \subref{fig:foto_velocity} illustrate that one value (75~km/h) can be member of more than one class.}
\label{fig:foto_fuzzy}
\end{figure}

Here, we briefly review the FOTO-method which allows a classification of 
locally measured traffic data according to the three-phase traffic theory.
The classification is based on a set of rules. 
It uses the aggregated data provided by a local detector (\ie velocity and flow) and decides 
to which traffic state the measured combination of observables most likely
belongs.
As the traffic phases S and J both denote phases of congested traffic, the 
distinction is not always obvious. 
Therefore, the set of rules employs a fuzzification of the input parameters as 
shown in figure~\ref{fig:foto_fuzzy}.

The fuzzification process transforms the measured value into fuzzy values which denote the 
degree of membership to the given classes. 
Hence, one value may belong to more than one class. 
This is illustrated in figure~\ref{fig:foto_velocity}, where fuzzificating the velocity of 75~km/h
shows that this velocity is both a ``medium'' and a ``high'' velocity. 
Yet the degree of membership to the class ``high'' is larger (0.75) than to the class~``medium''~(0.25). 

Based on this fuzzification of flow and velocity, the classification of traffic states via the FOTO-method works as 
follows~\cite{Kerner2004,KernerRehbornAleksicHaug2004}:
\begin{itemize}
	\item[\textbf{F1}] If the measured average velocity is classified as ``high'', 
			then the associated traffic phase is free flow (F)---independent of the flow rate's value.
	\item[\textbf{F2}] If the measured average velocity is a member of the class ``medium'', 
			the associated traffic phase corresponds to synchronized traffic (S).
	\item[\textbf{F3}] If the measured average velocity is ``low'' but the flow is ``high'', 
			the associated traffic phase is synchronized traffic (S).
	\item[\textbf{F4}] If both the measured average velocity and the flow are ``low'', the 
			associated traffic phase should be classified as a wide moving jam (J).
\end{itemize}
Due to non-exclusive memberships, a pair of flow and average velocity may match more than one of 
the above criteria. 
In this case, one has to chose the traffic state with the highest degree of membership of both velocity and flow. 

\section{The comfortable driving model (CDM)}
\label{app:cdm}
The CDM, originally called \textit{brake-light model}, 
is an advancement of the Nagel-Schreckenberg (NSM) cellular automaton with extensions for anticipatory driving.
Thereby, the CDM enables a vehicle to react more carefully to the preceding one. 
(In the NSM any preceding vehicle is ignored, unless a collision is imminent.) 
The model includes anticipatory effects by considering the status of the preceding vehicle's brake light $b_{n+1}$, 
which may be on (\ie $b_{n+1}(t)=1$) or off (\ie $b_{n+1}(t)=0$), by anticipating its own velocity 
$v_{\rm anti}(t)=\min(v_{n+1}(t),d_{n+1}(t))$, and by calculating the effective distance $d_n^{\rm eff}(t)$ as
\begin{equation}
\label{eqn:cdm_effective_distance} 
	d_{n}^\mathrm{eff}(t)=d_n(t)+\max\left(v_{\rm anti}(t)-d_{\rm safe},0\right)
\end{equation}
where the parameter $d_{\rm safe}$ governs the effectiveness of the anticipation.%

Vehicle motion results from the simultaneous application of several rules that are 
explained in the following:
\begin{enumerate}
	\item \textbf{Acceleration}:
	In the first step, a vehicle tries to accelerate to its maximum velocity. 
	To avoid unnecessary acceleration, it checks the status of its own and the preceding vehicle's brake light 
	and compares its time headway $t_\mathrm{h}(t)=d_n(t)/v_n(t)$ to a velocity-dependent
	interaction horizon $t_s(t)=\min\left(v_n(t),h\right)$.
	\begin{numcases}{v_n(t+1) \leftarrow}
		\min\left(v_n^{\max},v_n(t)+1\right), 	& if $b_n(t)=b_{n+1}(t)=0$ or $t_\mathrm{h}(t) \geq t_s(t)$,\nonumber\\
  		v_n(t),									& otherwise.
	\end{numcases}
	
	\item \textbf{Braking}:	Here, the vehicle checks whether it actually has to brake and updates the status of its brake light.
	The function $\Theta(\cdot)$ denotes the Heaviside step function. 
	\begin{eqnarray}
		v_n(t+1) & \leftarrow \min\left(d_n^{\rm eff}(t), v_n(t+1)\right)\\
		b_n(t+1) & \leftarrow 1 - \Theta\left(v_n(t+1) - v_n(t)\right)
	\end{eqnarray}
	
	\item \textbf{Determination of randomization parameter } \textbf{\textit{p}:}
	\begin{numcases}{p \leftarrow}
		\label{eqn:probabilistic_variable}
		p_b, & if $b_{n+1}(t)=1$ and $t_\mathrm{h}(t)<t_s(t)$,\nonumber\\
  		p_0, & if $v_n=0$,\\
  		p_d, & otherwise.\nonumber
	\end{numcases}

	\item \textbf{Dawdling}: In the CDM, this step influences both the vehicle's velocity and the 
	state of its brake light. Let $\xi$ be a (pseudo-)random number, uniformly distributed in $[0,1]$:

	\begin{numcases}{v_n(t+1) \leftarrow} 	
								\max\left(v_n(t+1)-1, 0\right), & if~$\xi < p$,\nonumber\\
								v_n(t+1), & otherwise.
	\end{numcases} 
	\begin{numcases}{b_n(t+1) \leftarrow}
								1, & if~$\xi < p~\textrm{and~}p=p_b$,\nonumber\\
								b_n(t+1), & otherwise.
	\end{numcases}

	\item \textbf{Vehicle motion}:
	\begin{equation}
		\label{eqn:cdm_move}
		x_n(t+1) \leftarrow x_n(t) + v_n(t+1)
	\end{equation}
\end{enumerate}
The variables are set to $p_d = 0.1$, $p_b = 0.94$, $p_0 = 0.5$, $h = 6$, and $d_{\rm safe} = 7$.

From the definition of the anticipated velocity $v_{\rm anti}(t)$ follows that in the CDM 
a vehicle's motion does not only dependent on the velocity of and the distance to the leading vehicle 
but also on the distance of the latter to its own predecessor.
This allows vehicles to accept time headways of below 1~s when driving 
at high velocities (cf. \cite{KnospeSantenSchadschneiderSchreckenberg2004,AppertRoland2009}). 
\end{appendix}

\section*{References}
\bibliographystyle{iopart-num}
\bibliography{references}

\end{document}